\documentclass[smallextended]{svjour3}       % onecolumn (second format)
\smartqed  % flush right qed marks, e.g. at end of proof
\usepackage[dvipdfmx]{graphicx}

\usepackage[T1]{fontenc}
\usepackage[latin9]{inputenc}
\usepackage{bm}
\usepackage{amstext}
\usepackage{latexsym,amsmath}
\usepackage{amssymb}
\usepackage[dvipdfmx]{hyperref}
\newcommand*\LyXbar{\rule[0.585ex]{1.2em}{0.25pt}}
 \journalname{}
\begin{document}

\title{Singular Behavior of the Macroscopic Quantity Near the Boundary for a
Lorentz-Gas Model with the Infinite-Range Potential}

%\subtitle{}

\titlerunning{Moment Singularity near the Boundary for the Infinite-Range Potential}        % if too long for running head

\author{Shigeru TAKATA         \and
        Masanari HATTORI %etc.
}


\institute{S. Takata \at
              Department of Aeronautics and Astronautics,
              Graduate School of Engineering,
              Kyoto University, Kyoto 615-8540, Japan;
              also at Research Project of Fluid Science and Engineering, 
              Advanced Engineering Research Center, Kyoto University, Kyoto 615-8540, Japan\\
%              Tel.: +123-45-678910\\
%             Fax: +123-45-678910\\
              \email{takata.shigeru.4a@kyoto-u.ac.jp}           %  \\
           \and
           M. Hattori \at
              Department of Aeronautics and Astronautics,
              Graduate School of Engineering,
              Kyoto University, Kyoto 615-8540, Japan;
              also at Research Project of Fluid Science and Engineering, 
              Advanced Engineering Research Center, Kyoto University, Kyoto 615-8540, Japan
}

\date{Received: date / Accepted: date}

\maketitle

\begin{abstract}
Possibility of the diverging gradient of the macroscopic quantity 
near the boundary is investigated by a mono-speed Lorentz-gas model, with a special attention to the regularizing effect of the grazing collision for the infinite-range potential on the velocity distribution function (VDF) and its influence on the macroscopic quantity. 
By careful numerical analyses of the steady one-dimensional boundary-value problem, 
it is confirmed that the grazing collision suppresses the occurrence
of a jump discontinuity of the VDF on the boundary. However, as the price for that regularization, the collision integral becomes no longer finite in the direction of the molecular velocity parallel to the boundary.
Consequently, the gradient of the macroscopic quantity diverges,
even stronger than the case of the finite-range potential.
A conjecture about the diverging rate in approaching the boundary is made as well for a wide range of the infinite-range potentials, accompanied by the numerical evidence.
\keywords{Kinetic theory of gases, Boltzmann equation, Infinite-range potential, Grazing collision, Lorentz gas, Kac model, Singularity}
\subclass{74A25 \and 76P05 \and 74G40}
\end{abstract}

\section{Introduction\label{sec:intro}}

It has been known for a long time that the velocity distribution function
(VDF) of molecules in a rarefied gas has a jump discontinuity, in general,
on the boundary in the direction of molecular velocity parallel to the
boundary, e.g. see Refs.~\cite{K69,S07}. Originating from this feature,
the macroscopic quantities defined as the moment of VDF change steeply
near the boundary in the direction normal to it. Here, the steep change
does not mean the Knudsen layer (the kinetic boundary layer) in slightly
rarefied gases, but rather means the singular behavior of those quantities
at the bottom of the ballistic non-equilibrium region with the thickness
of the mean-free-path of a molecule. The Knudsen layer is just an
example of such a non-equilibrium region. Note that the non-equilibrium
region extends much wider and possibly even to the entire region in
low pressure circumstances or in micro-scale physical systems. The
variation becomes steeper indefinitely in approaching the boundary,
and the variation rate diverges finally on the boundary. The diverging
rate follows a universality such that it depends on the local geometry
of the boundary. The detailed discussions can be found in Ref.~\cite{TT17}.

In the literature \cite{TF13,CLT14,TT17,TST19,CH15}, the diverging
rate has been discussed in the connection with a jump discontinuity
of the VDF both qualitatively and quantitatively. However, in those discussions
it is supposed that the collision integral can be split into the gain and
the loss term, namely the case where the collision frequency is finite.
This means that the investigated molecular models are the finite-range
potentials or the cutoff potentials if the infinite-range potentials
are in mind \cite{C88,S07}. The grazing
collisions that change the molecular trajectory only slightly have
been studied intensively for the infinite range potentials as an attractive mathematical topics
in the last two decades, e.g., Refs.~\cite{D95,DG00,V02,AV02,MS07,AMUXY10,AMUXY11,GS11,CH11,JL19},
and are found to have a regularizing effect on the VDF for such potentials. 

In view of those mathematical studies, it is expected that the jump
discontinuity of the VDF is not allowed even on the boundary for the
infinite-range potential, which may, in turn, suppress the diverging gradient 
 of macroscopic quantities because of the absence of its
origin. It motivates us to study whether
or not the diverging gradient occurs for the infinite range potentials
by using a mono-speed Lorentz-gas model equation. This model
equation, in place of the original Boltzmann equation, has already been used
in Ref.~\cite{T15} to investigate the propagation of the jump discontinuity
in the initial data and has been shown to capture the features of
the propagation well. In this sense, the present work may also be
regarded as an extension of Ref.~\cite{T15} to the steady one-dimensional
boundary-value problem. As will be clarified later, the grazing collisions
for the infinite range potential indeed do not allow the jump discontinuity
of the VDF on the boundary. Nevertheless, as the price for this regularizing
effect, the collision integral no longer remains finite;
consequently, the diverging gradient manifests itself more strongly than the case
of the finite range potential when approaching the boundary.

The paper is organized as follows. First, the mono-speed Lorentz-gas model
is introduced and the one-dimensional boundary-value problem is set
up in Sec.~\ref{sec:LorentzModel}.%
Thus, the singularity near the \textit{flat} boundary will be investigated.\footnote{Although the Lorentz-gas model will be considered in two-dimensional space both in the position and molecular velocity, the boundary that does not change its shape under a scale change will be called the \textit{flat} boundary, in place of the straight boundary, in the present paper.}
Then, in Sec.~\ref{sec:numerics},
two numerical methods are introduced. One is a rather direct approach
that is particularly suitable for the study of the finite-range and
the cutoff potential and is briefly explained in Sec.~\ref{subsec:directSolution}.
The other is the approach based on the Galerkin method, applicable
 to the infinite-range potential as well, and explained in detail in Sec.~\ref{subsec:GalerkinMethod}.
The numerical results are presented in Sec.~\ref{sec:results}.
The results for the cutoff potential with various cutoff sizes
and those for the corresponding infinite-range potential are compared
in the Maxwell-molecule-type case.
Furthermore, the diverging rate of
the gradient of the macroscopic quantity are identified
for the same case in Sec.~\ref{subsec:discussion}.
A conjecture on the diverging rate for other infinite range potentials
is made in Sec.~\ref{subsec:Conjecture}, accompanied by the additional numerical evidence. The paper is concluded in Sec.~\ref{sec:conclusion}.

\section{Lorentz-Gas Model\label{sec:LorentzModel}}

We consider the following mono-speed Lorentz-gas model that is two-dimensional
both in the position and the molecular velocity space in the present
paper. \begin{subequations}\label{toy1}
\begin{align}
\frac{\partial f}{\partial t}+\alpha_{i}\frac{\partial f}{\partial x_{i}} & =\int_{|\bm{\beta}|=1}b(|\bm{\alpha}\cdot\bm{\beta}|)\{f(t,\bm{x},\bm{\alpha}_{*})-f(t,\bm{x},\bm{\alpha})\}d\bm{\beta},\label{eq:toy1}\\
\bm{\alpha}_{*} & =\bm{\alpha}-2(\bm{\beta\cdot\bm{\alpha}})\bm{\beta}.\label{eq:alp1}
\end{align}
\end{subequations}The same model was used in Ref.~\cite{T15} for the
study of the grazing collision effects on the time evolution from
the initial data with a jump discontinuity. Here, $f$ is the dimensionless
velocity distribution function (VDF), $t$ is the dimensionless time,
$\bm{x}$ is the dimensionless position vector, and $\bm{\alpha}$,
$\bm{\alpha_{*}}$, and $\bm{\beta}$ are unit vectors, where the
reference scales of quantities are chosen in such a way that both
of the Strouhal and the Knudsen number are unity. The unit vectors
$\bm{\alpha}$ and $\bm{\alpha_{*}}$ represent the dimensionless
velocity of a molecule, the size of which does not change by the present
collision integral, i.e., the right-hand side. The molecular velocity
changes only its direction by the effect of the right-hand side. The
direction of change is represented by another unit vector $\bm{\beta}$.
The function $b$ represents the interaction effect and is non-negative.
Here, it is assumed to take the following form in order to mimic the
hard-disk and the inverse-power-law potential model:\footnote{The present definition of $b$ is different from that in Ref.~\cite{T15}
by the normalization factor.}\begin{subequations}\label{b1}
\begin{align}
b(|\bm{\alpha}\cdot\bm{\beta}|) & =B_{\gamma+2}^{-1}|\bm{\alpha}\cdot\bm{\beta}|^{\gamma},\quad(-3<\gamma\le1),\label{eq:b1}\\
B_{\mathrm{\gamma}} & =\int_{|\bm{\beta}|=1}|\bm{\alpha}\cdot\bm{\beta}|^{\gamma}d\bm{\beta}.\label{eq:nuM}
\end{align}
\end{subequations}As explained in Ref.~\cite{T15}, the setting $\gamma=1$
is the hard-disk potential, while the setting $\gamma=-\frac{n+1}{n-1}$
well mimics the angular singularity (or the grazing collision effect)
occurring in the Boltzmann equation for the $(n-1)$-th inverse-power-law
potential, where $n=5$ (or $\gamma=-3/2$) corresponds to the celebrated Maxwell molecule.
It should be noted that $B_{\gamma}$ is the (dimensionless) collision
frequency for the adopted interaction potential and remains finite
as far as $\gamma>-1$. The range $-1<\gamma<1$ is not
covered by the inverse-power-law potential and the collision integral
can be split into the so-called gain and loss term safely; this range
of $\gamma$ will be referred to the finite-range potential in the
present paper. For $-3<\gamma\le-1$ (or $n>2$), $B_{\gamma}$ is no longer finite
but diverges and the collision term can be treated only when the collision
integral is treated as a whole; this range of $\gamma$ will be referred
to the infinite-range potential in the present paper. The setting
$\gamma=-3$ (or $n=2$) corresponds to the Coulomb potential and the collision
term no longer remains finite. The factor $B_{\gamma+2}$ occurring
in (\ref{eq:b1}) is the effective collision frequency based on the
momentum change in collisions. As is seen from (\ref{eq:nuM}), it
does not diverge for $\gamma>-3$.

\subsection{Problem and Formulation}

In order to study the possibility of the diverging gradient of macroscopic
quantities, the following steady one-dimensional boundary-value problem
is considered for the above Lorentz-gas model \eqref{toy1}:\begin{subequations}\label{toy2}
\begin{align}
\alpha_{1}\frac{\partial f}{\partial x_{1}} & =\int_{|\bm{\beta}|=1}b(|\bm{\alpha}\cdot\bm{\beta}|)\{f(x_{1},\bm{\alpha}_{*})-f(x_{1},\bm{\alpha})\}d\bm{\beta},\label{eq:toy2}\\
\mbox{b.c. }f & =\frac{1}{2\pi}(1\pm c),\quad x_{1}=\mp\frac{1}{2},\ \alpha_{1}\gtrless0,\label{eq:bc2}
\end{align}
\end{subequations}where $0<c<1$ is a constant. The (dimensionless)
density $\rho$ is expressed as the following moment of $f$:\footnote{The $x_{1}$- and the $x_{2}$-component of the (dimensionless) mass
flow $\rho v_{1}$ and $\rho v_{2}$ are expressed as
\begin{equation*}
\rho v_{1} =\int_{|\bm{\alpha}|=1}\alpha_{1}fd\bm{\alpha},\quad
\rho v_{2} =\int_{|\bm{\alpha}|=1}\alpha_{2}fd\bm{\alpha}.
\end{equation*}
The $\rho v_{1}$ is constant because of the mass conservation law
obtained by the integration of (\ref{eq:toy2}) with respect to $\bm{\alpha}$.
As for $\rho v_{2}$, the similarity solution compatible to the problem in Sec.~\ref{subsec:symmetry}
leads to $\rho v_{2}\equiv0$.
Hence, our primary target is to study the behavior of $\rho$ near
the boundaries $x_{1}=\pm1/2$.}
\begin{equation}
\rho=\int_{|\bm{\alpha}|=1}fd\bm{\alpha},\label{eq:dens1}
\end{equation}
the behavior of which near the boundary $x_{1}=-1/2$ is the primary
target of the present study. 

By noting the relation
\begin{equation}
|\bm{\alpha}\cdot\bm{\beta}|=\Big(\frac{1-\bm{\alpha}\cdot\bm{\alpha}_{*}}{2}\Big)^{1/2},\label{eq:inpro}
\end{equation}
the above problem \eqref{toy2} is reduced to that for $g\equiv(2\pi f-1)/c$:\begin{subequations}\label{toy3}
\begin{align}
\sin\theta\frac{\partial g}{\partial x_{1}} & =C_{\gamma}[g],\label{eq:toy3}\\
g & =\pm1,\quad x_{1}=\mp\frac{1}{2},\ \sin\theta\gtrless0.\label{eq:bc3}
\end{align}
\end{subequations}Here
\begin{align}
C_{\gamma}[g] & =\frac{1}{B_{\gamma+2}}\int_{-\pi}^{\pi}\Big(\frac{1-\cos\theta_{*}}{2}\Big)^{\gamma/2}\{g(x_{1},\theta+\theta_{*})-g(x_{1},\theta)\}d\theta_{*}\nonumber \\
 & =\frac{1}{B_{\gamma+2}}\int_{-\pi}^{\pi}|\sin\frac{\phi-\theta}{2}|^{\gamma}\{g(x_{1},\phi)-g(x_{1},\theta)\}d\phi,\displaybreak[0]\label{eq:col3}\\
B_{\gamma} & \equiv\int_{|\bm{\beta}|=1}|\bm{\alpha}\cdot\bm{\beta}|^{\gamma}d\bm{\beta}=\int_{-\pi}^{\pi}|\cos\varphi|^{\gamma}d\varphi=2\int_{0}^{\pi}|\sin\frac{\phi}{2}|^{\gamma}d\phi,
\end{align}
and $\theta$ and $\theta+\theta_{*}$ respectively indicate the clockwise
angle of the unit vectors $\bm{\alpha}$ and $\bm{\alpha}_{*}$ measured
from the $x_{2}$-direction. Note that in (\ref{eq:col3}), the range
of integration for $\phi$ is shifted by $\theta$ because of the
$2\pi$-periodicity. The density is then reduced to the following
moment of $g$:
\begin{equation}
\rho(x_{1})=1+\frac{c}{2\pi}\int_{-\pi}^{\pi}g(x_{1},\theta)d\theta\equiv1+c\rho_{g}(x_{1}).\label{eq:dens2}
\end{equation}

\subsection{Angular Cutoff}

When $-1<\gamma$, the $C_{\gamma}$ defined in (\ref{eq:col3}) can
be treated separately as:\begin{subequations}\label{noncut}
\begin{align}
C_{\gamma}[g] & =C_{\gamma}^{+}[g]-\nu_{\gamma}g,\displaybreak[0]\label{eq:noncut}\\
C_{\gamma}^{+}[g] & =\int_{-\pi}^{\pi}b_{\gamma}(\phi-\theta)g(x_{1},\phi)d\phi,\displaybreak[0]\label{eq:gain}\\
\nu_{\gamma} & =\int_{-\pi}^{\pi}b_{\gamma}(\phi-\theta)d\phi=\int_{-\pi}^{\pi}b_{\gamma}(\phi)d\phi,\displaybreak[0]\label{eq:loss}\\
 & b_{\gamma}(\varphi)\equiv\frac{1}{B_{\gamma+2}}|\sin\frac{\varphi}{2}|^{\gamma}.\label{eq:bg}
\end{align}
\end{subequations}It is not the case, however, when $-3<\gamma\le-1$,
since $b_{\gamma}(\varphi)$ is singular for $\varphi\to0$ strongly
enough for the integrability not to be assured. Physically, it implies
that the grazing events that are little effective to change the particle
velocity are all counted as the collision. Hence in the literature,
the truncation of the range of $\varphi$, the so-called angular cutoff
\cite{C88}, is introduced in order to avoid counting such an enormous
amount of grazing events. The infinite-range potential with the cutoff
will be simply called the cutoff potential in what follows. With the
size of the cutoff $\epsilon$, the following notations for the cutoff
potential are introduced here:\begin{subequations}\label{cutoff}
\begin{align}
C_{\gamma,\epsilon}[g] & =C_{\gamma,\epsilon}^{+}[g]-\nu_{\gamma,\epsilon}g,\displaybreak[0]\label{eq:cutoff}\\
C_{\gamma,\epsilon}^{+}[g] & =\int_{-\pi}^{\pi}b_{\gamma,\epsilon}(\phi-\theta)g(x_{1},\phi)d\phi,\displaybreak[0]\label{eq:gaincut}\\
\nu_{\gamma,\epsilon} & =\int_{-\pi}^{\pi}b_{\gamma,\epsilon}(\phi-\theta)d\phi=\int_{-\pi}^{\pi}b_{\gamma,\epsilon}(\phi)d\phi,\label{eq:losscut}
\end{align}
\end{subequations}where
\begin{align}
b_{\gamma,\epsilon}(\varphi) & =\begin{cases}
\frac{B_{\gamma+2}}{B_{\gamma+2,\epsilon}}b_{\gamma}(\varphi), & \epsilon<\varphi<2\pi-\epsilon\\
0, & \mathrm{otherwise}
\end{cases},\quad(0<\varphi<2\pi),\displaybreak[0]\label{eq:bgcut}\\
B_{\gamma,\epsilon} & =2\int_{\epsilon}^{\pi}|\sin\frac{\phi}{2}|^{\gamma}d\phi,
\end{align}
and the factor $B_{\gamma+2}/B_{\gamma+2,\epsilon}$ is used so that
the effective collision cross-section based on the momentum change
\cite{B94,T15} becomes common between the cutoff and the infinite-range
potential for the same $\gamma$.

\subsection{Small Reduction Using Problem Symmetry \label{subsec:symmetry}}

The $g$ having the following symmetry matches the boundary-value
problem \eqref{toy3}:\begin{subequations}\label{sym}
\begin{align}
g(\cdot,\theta) & =g(\cdot,\pi-\theta),\quad(\frac{\pi}{2}<\theta<\pi),\label{eq:sym1}\\
g(\cdot,\theta) & =g(\cdot,-\pi-\theta),\quad(-\pi<\theta<-\frac{\pi}{2}),\label{eq:sym2}\\
g(x_{1},\theta) & =-g(-x_{1},-\theta),\quad(0<x_{1}<\frac{1}{2},\ -\frac{\pi}{2}<\theta<\frac{\pi}{2}).\label{eq:sym3}
\end{align}
\end{subequations}The properties (\ref{eq:sym1}) and (\ref{eq:sym2})
admit the following expression of $\rho_{g}$
\begin{equation}
\rho_{g}(x_{1})=\frac{1}{\pi}\int_{-\pi/2}^{\pi/2}g(x_{1},\theta)d\theta,\label{eq:rho_ghalf}
\end{equation}
and the following transformation of $C_{\gamma}$:
\begin{align}
%\begin{equation*}
C_{\gamma}[g] 
& =\int_{-\pi}^{\pi}b_{\gamma}(\phi-\theta)\{g(x_{1},\phi)-g(x_{1},\theta)\}d\phi\displaybreak[0]\nonumber \\
& =\int_{-\pi/2}^{\pi/2}b_{\gamma}(\phi-\theta)\{g(x_{1},\phi)-g(x_{1},\theta)\}d\phi\nonumber \\
& +\int_{-\pi}^{-\pi/2}b_{\gamma}(\phi-\theta)\{g(x_{1},\phi)-g(x_{1},\theta)\}d\phi\nonumber \\
& +\int_{\pi/2}^{\pi}b_{\gamma}(\phi-\theta)\{g(x_{1},\phi)-g(x_{1},\theta)\}d\phi\displaybreak[0]\nonumber \\
& =\int_{-\pi/2}^{\pi/2}b_{\gamma}(\phi-\theta)\{g(x_{1},\phi)-g(x_{1},\theta)\}d\phi\nonumber \\
& +\int_{-\pi}^{-\pi/2}b_{\gamma}(\phi-\theta)\{g(x_{1},-\pi-\phi)-g(x_{1},\theta)\}d\phi\nonumber \\
& +\int_{\pi/2}^{\pi}b_{\gamma}(\phi-\theta)\{g(x_{1},\pi-\phi)-g(x_{1},\theta)\}d\phi\displaybreak[0]\nonumber \\
& =\int_{-\pi/2}^{\pi/2}b_{\gamma}(\phi-\theta)\{g(x_{1},\phi)-g(x_{1},\theta)\}d\phi\nonumber \\
& +\int_{-\pi/2}^{0}b_{\gamma}(-\psi-\pi-\theta)\{g(x_{1},\psi)-g(x_{1},\theta)\}d\psi\quad(\psi=-\pi-\phi)\nonumber \\
& +\int_{0}^{\pi/2}b_{\gamma}(\pi-\psi-\theta)\{g(x_{1},\psi)-g(x_{1},\theta)\}d\psi\quad(\psi=\pi-\phi)\displaybreak[0]\nonumber \\
&
 =\int_{-\pi/2}^{\pi/2}\{b_{\gamma}(\phi-\theta)+b_{\gamma}(\pi-\phi-\theta)\}\{g(x_{1},\phi)-g(x_{1},\theta)\}d\phi,\label{eq:coltrans}
\end{align}
%\end{equation*}
where the relation
\begin{equation}
b_{\gamma}(-\psi-\theta-\pi)=b_{\gamma}(\psi+\theta+\pi)=b_{\gamma}(\psi+\theta-\pi)=b_{\gamma}(\pi-\psi-\theta),\label{eq:bgprop}
\end{equation}
has been used. Moreover, by using (\ref{eq:sym3}), the problem \eqref{toy3}
is reduced to the following problem in $-1/2<x_{1}<0$ and $-\pi/2<\theta<\pi/2$:\begin{subequations}\label{prob}
\begin{align}
\sin\theta\frac{\partial g}{\partial x_{1}} & =C_{\gamma}[g],\label{eq:prob}\\
C_{\gamma}[g] & =\int_{-\pi/2}^{\pi/2}\{b_{\gamma}(\phi-\theta)+b_{\gamma}(\pi-\phi-\theta)\}\{g(x_{1},\phi)-g(x_{1},\theta)\}d\phi,\label{eq:probcol}\\
 & \mathrm{b.c.\ }\begin{cases}
g(0,\theta)=-g(0,-\theta) & -\pi/2<\theta<0\\
g(-1/2,\theta)=1 & 0<\theta<\pi/2
\end{cases}.\label{eq:probbc}
\end{align}
\end{subequations}Since $b_{\gamma,\epsilon}$ matches the same relation
(\ref{eq:bgprop}) as $b_{\gamma}$, the problem \eqref{prob} is
written for the corresponding cutoff potential by simply replacing
$C_{\gamma}$ with $C_{\gamma,\epsilon}$:\begin{subequations}\label{probcut}
\begin{align}
\sin\theta\frac{\partial g}{\partial x_{1}} & =C_{\gamma,\epsilon}[g],\label{eq:probcut}\\
C_{\gamma,\epsilon}[g] & =\int_{-\pi/2}^{\pi/2}\{b_{\gamma,\epsilon}(\phi-\theta)+b_{\gamma,\epsilon}(\pi-\phi-\theta)\}\{g(x_{1},\phi)-g(x_{1},\theta)\}d\phi,\label{eq:probcutcol}\\
 & \mathrm{b.c.\ }\begin{cases}
g(0,\theta)=-g(0,-\theta) & -\pi/2<\theta<0\\
g(-1/2,\theta)=1 & 0<\theta<\pi/2
\end{cases}.\label{eq:probcutbc}
\end{align}
\end{subequations}Remind that $C_{\gamma}$ can be treated as\begin{subequations}\label{probcol1}
\begin{align}
C_{\gamma}[g] & =C_{\gamma}^{+}[g]-\nu_{\gamma}g,\label{eq:probcol1}\\
C_{\gamma}^{+}[g] & =\int_{-\pi/2}^{\pi/2}\{b_{\gamma}(\phi-\theta)+b_{\gamma}(\pi-\phi-\theta)\}g(x_{1},\phi)d\phi,\label{eq:probgain1}
\end{align}
\end{subequations}only when $-1<\gamma$. When $-3<\gamma\le-1$, it is
$C_{\gamma,\epsilon}$ which can be treated separately:\begin{subequations}\label{probcol2}
\begin{align}
C_{\gamma,\epsilon}[g] & =C_{\gamma,\epsilon}^{+}[g]-\nu_{\gamma,\epsilon}g,\label{eq:probcol2}\\
C_{\gamma,\epsilon}^{+}[g] & =\int_{-\pi/2}^{\pi/2}\{b_{\gamma,\epsilon}(\phi-\theta)+b_{\gamma,\epsilon}(\pi-\phi-\theta)\}g(x_{1},\phi)d\phi.\label{eq:probgain2}
\end{align}
\end{subequations}

\section{Methods of Numerical Analyses\label{sec:numerics}}

In numerically treating the problem formulated in Sec.~\ref{subsec:symmetry},
the grid points in $\theta$-space are arranged to be symmetric with
respect to $\theta=0$ in the region $-\pi/2\le\theta\le\pi/2$ so
as to make $2N$ small intervals in both the positive and negative
side:
\begin{equation*}
0 =\theta^{(0)}<\theta^{(1)}<\cdots<\theta^{(2N-1)}<\theta^{(2N)}=\frac{\pi}{2},
\ \theta^{(-j)}=-\theta^{(j)},\ (j=1,\dots,2N).
\end{equation*}
Two different methods are prepared. One is the method making use of
the numerical kernel \cite{SOA89} as in Ref.~\cite{HT15} and is referred
to as the direct method in the present paper. The direct method is
able to treat \eqref{probcol1} and \eqref{probcol2} without numerical
(or unphysical) oscillation, even when $g$ has a jump discontinuity
at $\theta=0$ on the boundary; see Appendix~\ref{sec:kernel}. This is the
primary merit of the method and makes it suitable for the finite-range
and the cutoff potential cases. As will be observed later through
the results for the infinite-range potential, the jump discontinuity
tends to vanish as $\epsilon\to0$, but the collision integral instead
tends to diverge at $\theta=0$ on the boundary, i.e., in the direction
parallel to the boundary. This implies that a weaker formulation is
unavoidable to study the infinite-range potential and motivates the
approach using the Galerkin method.

Since the jump discontinuity is expected not to appear for the infinite-range
potential, $g(x_{1},\theta)$ is approximated by the set of quadratic basis
functions continuously. Then, the problem \eqref{prob}
is discretized by the projection into the space of the same basis
functions in the Galerkin method. If the same is applied to the cutoff
potential \eqref{probcut}, artificial oscillations occur due to the
jump discontinuity. However, as will be observed later in Sec.~\ref{sec:results},
it affects little to the behavior of $\rho_{g}$.

\subsection{Direct Method for the Finite-Range and the Cutoff Potential \label{subsec:directSolution}}

For the finite-range and the cutoff potential, the collision integral
can be split into the loss and the gain term safely, and the problem
is formally solved as :
\begin{align*}
g(x_{1},\theta) & =\begin{cases}
{\displaystyle e^{-\frac{\nu}{\sin\theta}(x_{1}+\frac{1}{2})}+\int_{-1/2}^{x_{1}}\frac{1}{\sin\theta}C^{+}[g]{(s,\theta)}e^{-\frac{\nu}{\sin\theta}(x_{1}-s)}ds,} \\
\qquad\qquad\qquad\qquad (0<\theta<\pi/2,\ -1/2<x_{1}<0),\\
{\displaystyle -g(0,-\theta)e^{-\frac{\nu}{\sin\theta}x_{1}}+\int_{0}^{x_{1}}\frac{1}{\sin\theta}C^{+}[g]{(s,\theta)}e^{-\frac{\nu}{\sin\theta}(x_{1}-s)}ds,} \\
\qquad\qquad\qquad\qquad (-\pi/2<\theta<0,\ -1/2<x_{1}<0),
\end{cases}
\end{align*}
where the subscript $\gamma$ and $\epsilon$ are omitted in $C^{+}$
and $\nu$, since there is no need of discrimination in the present
context, while the arguments of $C^{+}[g]$ are indicated explicitly for clarity.
The same omission convention will be applied in what follows, 
unless any confusions/ambiguities are expected. 
In the direct method, the solution
$g$ is constructed by iteration from its initial guess. In this process,
by substituting $g$ of the following approximation (the expansion
in terms of the quadratic basis functions $\{Y_{j}(\theta)\}$,
see Appendix~\ref{sec:kernel}):
\begin{equation}
g(x_{1},\theta)=\sum_{j=-2N}^{2N}g_{j}(x_{1})Y_{j}(\theta),\label{eq:basisFunc}
\end{equation}
with $g_{j\gtrless0}$ being the value on the grid points in $\theta\gtrless0$,
the $C^{+}[g]$ at the grid point $\theta=\theta^{(i)}$ is expressed
as $C^{+}[g]{(s,\theta^{(i)})}=\sum_{j=-2N}^{2N}g_{j}(s)C^{+}[Y_{j}]{(\theta^{(i)})}$.
In this expression, the discrimination between $j=\pm0$ is made when
there is a jump discontinuity of $g$ at $\theta=0$. Although the
analytical expression of $C^{+}[Y_{j}]$ in $\theta$ is obtained
{[}see Appendix~\ref{sec:kernel}, especially the descriptions below (\ref{eq:disc_g})
there, for more details{]}, $C^{+}[Y_{j}]{(\theta^{(i)})}$ at the
grid point of $\theta$ are stored beforehand as the numerical kernel
\cite{SOA89} in order to avoid repeating the same computation in
the iteration in solving $g$. The integration with respect to $s$
is performed analytically after the quadratic interpolation
of the data $C^{+}[g]{(s,\theta)}$ at the discrete position in $s$.

\subsection{Galerkin Method \label{subsec:GalerkinMethod}}

In the Galerkin method \cite{KR19}, irrespective of whether or not $g$ has a
jump discontinuity in $\theta$, the approximate expression (\ref{eq:basisFunc})
of $g$ in terms of the quadratic basis functions $\{Y_{j}(\theta)\}$
is used as it stands, i.e., without the discrimination between $j=\pm0$.
More precisely, even for the finite-range and the cutoff potential,
where $g$ is expected to have a jump discontinuity, (\ref{eq:basisFunc})
with $g_{0}=g_{-0}$ will be used in the solution process (see Appendix~\ref{sec:kernel}).
Before going into details, it should be noted that, thanks to the
symmetric arrangement of the grid points $\theta^{(j)}$, it holds
that $Y_{j}(\theta)=Y_{-j}(-\theta)$ and that $g_{j}(x_{1})=-g_{-j}(-x_{1})$
from the property (\ref{eq:sym3}). 

In order to construct the numerical procedure by the Galerkin method,
first substitute (\ref{eq:basisFunc}) into (\ref{eq:prob}) with
(\ref{eq:probcol}) and then integrate the result multiplied with
$Y_{i}(\theta)\ (i=-2N,\dots,2N)$ with respect to $\theta$. The
result is that
\begin{align}
\sum_{j=-2N}^{2N}A_{i,j}\frac{dg_{j}}{dx_{1}} & =\sum_{j=-2N}^{2N}D_{i,j}g_{j},\label{eq:GL1}\\
A_{i,j} & =\int_{-\pi/2}^{\pi/2}Y_{i}(\theta)Y_{j}(\theta)\sin\theta\ d\theta,\nonumber \\
D_{i,j} & =\int_{-\pi/2}^{\pi/2}Y_{i}(\theta)C[Y_{j}]{(\theta)}d\theta.\nonumber 
\end{align}
Here again the subscripts $\gamma$ and $\epsilon$ to $C$ are omitted.
Note that the integration in the definitions of $A_{i,j}$ and $D_{i,j}$
can be done analytically and that \begin{subequations}\label{eq:A_prop}
\begin{align}
 & A_{i,j}=A_{j,i}=-A_{-i,-j},\quad(i,j=0,\pm1,\dots,\pm2N),\\
 & A_{-i,j}=A_{i,-j}=0,\quad(i,j=1,\dots,2N),
\end{align}
\end{subequations}by definition. Moreover, $D_{i,j}=D_{j,i}$ holds,
since $C$ is self-adjoint. In order to solve (\ref{eq:GL1}), it
is relevant to check whether or not the following $(4N+1)\times(4N+1)$-symmetric
matrix $A$ is regular:
\begin{align*}
A\equiv[A_{i,j}]= & \left[\begin{array}{ccccccc}
A_{-2N,-2N} & \cdots & A_{-2N,-1} & A_{-2N,0} & A_{-2N,1} & \cdots & A_{-2N,2N}\\
\vdots & \ddots & \vdots & \vdots & \vdots & \ddots & \vdots\\
A_{-1,-2N} & \cdots & A_{-1,-1} & A_{-1,0} & A_{-1,1} & \cdots & A_{-1,2N}\\
A_{0,-2N} & \cdots & A_{0,-1} & A_{0,0} & A_{0,1} & \cdots & A_{0,2N}\\
A_{1,-2N} & \cdots & A_{1,-1} & A_{1,0} & A_{1,1} & \cdots & A_{1,2N}\\
\vdots & \ddots & \vdots & \vdots & \vdots & \ddots & \vdots\\
A_{2N,-2N} & \cdots & A_{2N,-1} & A_{2N,0} & A_{2N,1} & \cdots & A_{2N,2N}
\end{array}\right].
\end{align*}
By direct calculations, it was observed that $A$ is not full rank
and is rank deficient by one when $N=1,2,3$. This strongly suggests
that the rank deficiency is due to the factor $\sin\theta$ in front
of the spatial derivative of $g$ in (\ref{eq:prob}) is zero and
the differential equation degenerates at $\theta=0$; thus the same
rank deficiency is expected for other integer $N$. In what follows,
the numerical procedure is constructed by supposing $\mathrm{rank}A=4N$,
i.e., the rank deficiency by one.

Thanks to the property \eqref{eq:A_prop}, the matrix $A$ is expressed
as
\begin{align}
A=[A_{i,j}]= & \begin{array}{c}
\left[\begin{array}{c|c}
A^{(-)} & 0\\
\hline 0 & A^{(+)}
\end{array}\right]\begin{array}{c}
(i<0)\\
(i>0)
\end{array}\end{array},\label{eq:Amatrix}\\
 & \begin{array}{cc}
(j<0),(j>0)\end{array}\nonumber 
\end{align}
where $\text{\LyXbar\text{\LyXbar}}$ indicates the row $i=0$, $\Big|$
indicates the column $j=0$, and $A^{(+)}$ and $A^{(-)}$ are $2N\times2N$
symmetric matrices. The row $i=0$ and the column $j=0$ vector are
non-zero. From \eqref{eq:A_prop}, it is clear that $A_{0,0}=0$ and
that $A_{i,j}^{(+)}=-A_{-i,-j}^{(-)}$, implying that both $A^{(+)}$
and $A^{(-)}$ are regular under the assumption $\mathrm{rank}A=4N$.
Consequently, it follows that the $j>0$ part of the row $i=0$ of
$A$ is expressed by the linear combination of the row vectors of
$A^{(+)}$, while the $j<0$ part of the same row is expressed by
the linear combination of the row vectors of $A^{(-)}$. That is,
there are two sets of constants $\{c_{1},\dots c_{2N}\}$ and $\{\tilde{c}_{1},\dots\tilde{c}_{2N}\}$
such that
\begin{align*}
A_{0,j} & =\sum_{i=1}^{2N}c_{i}A_{i,j}^{(+)},\quad(j>0),\\
A_{0,-j} & =\sum_{i=1}^{2N}\tilde{c}_{i}A_{-i,-j}^{(-)}=-\sum_{i=1}^{2N}\tilde{c}_{i}A_{i,j}^{(+)},\quad(j>0).
\end{align*}
Since $A_{0,-j}=-A_{0,j}$ by \eqref{eq:A_prop}, it follows that
$c_{i}=\tilde{c}_{i}$ and that 
\[
c_{i}=\sum_{j=1}^{2N}A_{0,j}A_{j,i}^{(+)-1},\quad(i=1,2,\dots2N).
\]
By the same $c_{i}$, it is seen that

\[
\sum_{i=1}^{2N}c_{i}(A_{i,0}+A_{-i,0})=\sum_{i=1}^{2N}c_{i}(A_{i,0}-A_{i,0})=0=A_{0,0}.
\]
Hence, the row vector $i=0$ of $A$ is recovered by the linear combination
of the other row vectors with coefficients $\{c_{i}\}$.

Now, on one hand, (\ref{eq:GL1}) with $i=0$ is
\[
\sum_{j=-2N}^{2N}A_{0,j}\frac{dg_{j}}{dx_{1}}=\sum_{j=-2N}^{2N}D_{0,j}g_{j},
\]
while, on the other hand, since $A_{0,j}$ is expressed by the combination
of the other rows, the left-hand side is rewritten as
\begin{align*}
\mbox{L.H.S.}= & \sum_{j=-2N}^{2N}\{\sum_{i=1}^{2N}c_{i}(A_{i,j}+A_{-i,j})\}\frac{dg_{j}}{dx_{1}}\\
= & \sum_{j=-2N}^{2N}\{\sum_{i=1}^{2N}c_{i}(D_{i,j}+D_{-i,j})\}g_{j}.
\end{align*}
Hence, it follows that
\[
\sum_{j=-2N}^{2N}\big\{ D_{0,j}-\sum_{i=1}^{2N}c_{i}(D_{i,j}+D_{-i,j})\big\} g_{j}=0.
\]
By putting
\[
\widetilde{D}_{j}\equiv D_{0,j}-\sum_{i=1}^{2N}c_{i}(D_{i,j}+D_{-i,j}),\quad(j=-2N,\dots2N),
\]
and further assuming $\widetilde{D}_{0}\ne0$,\footnote{The validity of this assumption should be checked numerically. It is reasonable, however, since the derivative term degenerates
when $\theta=0$ and the solution for $\theta=0$ is determined solely
by the collision term $C[g]$. Equation~(\ref{eq:g0}) below is its
reflection. } $g_{0}$ is expressed as 
\begin{equation}
g_{0}=-\frac{1}{\widetilde{D}_{0}}\sum_{j=1}^{2N}(\widetilde{D}_{j}g_{j}+\widetilde{D}_{-j}g_{-j}).\label{eq:g0}
\end{equation}
After the preparation above, the problem is reduced to solving the
following problem for $g_{j}$ with $j\ne0$ that is obtained by the
substitution of (\ref{eq:g0}) to (\ref{eq:GL1}):
\begin{align*}
\sum_{j\ne0}\overline{A}_{i,j}\frac{dg_{j}}{dx_{1}} & =\sum_{j\ne0}\overline{D}_{i,j}g_{j},\quad(i=\pm1,\dots\pm2N),\\
\overline{A}_{i,j} & =A_{i,j}-\frac{A_{i,0}}{\widetilde{D}_{0}}\widetilde{D}_{j},\quad
\overline{D}_{i,j}  =D_{i,j}-\frac{D_{i,0}}{\widetilde{D}_{0}}\widetilde{D}_{j}.
\end{align*}
It should be noted that, although $A$ without $i=0$ row and without
$j=0$ column is regular, it is not clear whether or not $\overline{A}=[\overline{A}_{ij}]$
is regular. Nevertheless, it is natural to suppose that $\overline{A}$
is regular. Then, the problem is further reduced to%
\begin{align}
\frac{dg_{i}}{dx_{1}} & =\sum_{j\ne0}\mathcal{D}_{i,j}g_{j},\quad(i=\pm1,\dots\pm2N),\label{eq:GL2}\\
\mathcal{D}_{i,j} & =\sum_{k\ne0}\overline{A}_{i,k}^{-1}\overline{D}_{k,j},\quad(i,j=\pm1,\dots\pm2N),\nonumber 
\end{align}%
and to the eigenvalue problem of the new $4N\times4N$ matrix $\mathcal{D}=[\mathcal{D}_{i,j}]$.

In order to study the eigenvalue problem of $\mathcal{D}$, first
go back to the $(4N+1)\times(4N+1)$ matrix $D\equiv[D_{i,j}]$, which is a discrete version
of $C[g]$. Then, the $(4N+1)$-dimensional unit vector $^{t}\!(1,\dots,1)/\sqrt{4N+1}$
corresponding to $g=1/\pi$, the null (or the collision invariant)
of $C$, is the eigenvector for the eigenvalue zero of $D$, where the superscript
$t$ indicates the transpose of the row vector. 
Then, since $\overline{D}_{i,0}=0$ by construction, 
it holds that $\overline{D}\bm{u}_{(0)}=\bm{0}$ and $\mathcal{D}\bm{u}_{(0)}=\bm{0}$,
where $\overline{D}\equiv[\overline{D}_{i,j}]$ ($i,j=\pm1,\dots,\pm 2N$) is a $4N\times4N$ matrix
and $\bm{u}_{(0)}\equiv\,^{t}\!(1,\dots,1)/\sqrt{4N}$ is a $4N$-dimensional unit vector. 
That is, $\bm{u}_{(0)}$ is the eigenvector for the eigenvalue zero of
$\overline{D}$ and $\mathcal{D}$. Moreover, thanks to the
symmetric arrangement of grid points in $\theta$-space and the symmetry
of the problem (\ref{eq:sym3}), the eigenvalues of $\mathcal{D}$
appear pairwise in the sense that if $\lambda$ is a nonzero-eigenvalue,
$-\lambda$ is also the eigenvalue and the eigenvector for $\lambda$
and that for $-\lambda$ have the reversed order of components each
other. Because of the pairwise occurrence of nonzero eigenvalues,
the eigenvalue zero ought to be multiple, since $\mathcal{D}$ has
$4N$ eigenvalues. Now assume that the multiplicity of the eigenvalue
zero is two and denote the other $4N-2$ non-zero eigenvalues by $\text{\ensuremath{\lambda_{q}}}$
($q=\pm1,\pm2,\dots,\pm2N\mp1$), where it is set that $\lambda_{q}=-\lambda_{-q}$
and the real part of $\lambda_{q>0}$ is positive. Further supposing
that $\lambda_{q}\ne\lambda_{p}$ for $q\ne p$,\footnote{The property $\lambda_{q}\ne\lambda_{p}$ for $q\ne p$ has been confirmed
numerically. It has also been found that $\lambda$'s are all real,
though they are not obvious beforehand.} the unit eigenvectors $\bm{u}_{(q)}$ for the eigenvalue $\lambda_{q}$
form the basis of the $4N$-dimensional vector space together with
the unit eigenvector $\bm{u}_{(0)}$ and the unit generalized eigenvector
$\bm{u}_{*}$ for the eigenvalue zero, where $\bm{u}_{*}$ is chosen
to be perpendicular to $\bm{u}_{(0)}$ for the later convenience.
Then, making the matrix $P$ as 
\[
P=[\bm{u}_{(-2N+1)},\dots,\bm{u}_{(-1)},\bm{u}_{(1)},\dots,\bm{u}_{(2N-1)},\bm{u}_{(0)},\bm{u}_{*}]
\]
and multiplying its inverse $P^{-1}$ with (\ref{eq:GL2}) from the
left result in
\begin{equation}
\frac{dz_{i}}{dx_{1}}=\sum_{j\ne0}\mathcal{M}_{i,j}z_{j},\quad(i=\pm1,\dots\pm2N),\label{eq:GL2-1}
\end{equation}
where
\begin{align}
\bm{z} & =P^{-1}\bm{g},\\
\mathcal{M} & =P^{-1}\mathcal{D}P %\nonumber \\
% & 
=\left[\begin{array}{cccccccc}
\lambda_{-2N+1}\\
 & \ddots &  &  &  & 0\\
 &  & \lambda_{-1}\\
 &  &  & \lambda_{1}\\
 &  &  &  & \ddots\\
 & 0 &  &  &  & \lambda_{2N-1}\\
 &  &  &  &  &  & 0 & \xi\\
 &  &  &  &  &  & 0 & 0
\end{array}\right],
\end{align}
with $\xi$ being a certain constant. It is easy to solve (\ref{eq:GL2-1}) as
\begin{align*}
\bm{z} & =\left[\begin{array}{c}
\eta_{-2N+1}e^{\lambda_{-2N+1}(x_{1}+\frac{1}{2})}\\
\vdots\\
\eta_{-1}e^{\lambda_{-1}(x_{1}+\frac{1}{2})}\\
\eta_{1}e^{\lambda_{1}(x_{1}-\frac{1}{2})}\\
\vdots\\
\eta_{2N-1}e^{\lambda_{2N-1}(x_{1}-\frac{1}{2})}\\
\eta_{*}\xi x_{1}+\eta_{0}\\
\eta_{*}
\end{array}\right],
\end{align*}
and thus $\bm{g}(=P\bm{z})$ is obtained in the form:
\begin{align}
\bm{g}(x_{1}) & =\sum_{q=1}^{2N-1}\{\eta_{q}e^{\lambda_{q}(x_{1}-\frac{1}{2})}\bm{u}_{(q)}+\eta_{-q}e^{\lambda_{-q}(x_{1}+\frac{1}{2})}\bm{u}_{(-q)}\} \nonumber\\
&\qquad\qquad+(\eta_{*}\xi x_{1}+\eta_{0})\bm{u}_{(0)}+\eta_{*}\bm{u}_{*}\nonumber \\
 & =\sum_{q=1}^{2N-1}\{\eta_{q}e^{\lambda_{q}(x_{1}-\frac{1}{2})}\bm{u}_{(q)} +\eta_{-q}e^{-\lambda_{q}(x_{1}+\frac{1}{2})}\bm{u}_{(-q)}\}\nonumber\\
&\qquad\qquad+(\eta_{*}\xi x_{1}+\eta_{0})\bm{u}_{(0)}+\eta_{*}\bm{u}_{*},\label{eq:g_dc}
\end{align}
where $\lambda_q=-\lambda_{-q}$ has been used.
The $\eta_{0},\eta_{*},\eta_{\pm1},\dots,\eta_{\pm(2N-1)}$ are
unknown constants and will be determined by using the
conditions at $x_{1}=0$ and $x_{1}=-1/2$.

Consider first the condition at $x_{1}=0$. Let $\bm{a}^{-}$ be $\bm{a}$
with its component order reversed. Then, the condition at $x_{1}=0$
is written as $\bm{g}(0)=-\bm{g}^{-}(0)$. In the meantime, the expression
(\ref{eq:g_dc}) yields
\begin{align}
\bm{g}(0) & =\sum_{q=1}^{2N-1}\eta_{q}e^{-\lambda_{q}/2}\bm{u}_{(q)}+\sum_{q=1}^{2N-1}\eta_{-q}e^{-\lambda_{q}/2}\bm{u}_{(-q)}+\eta_{0}\bm{u}_{(0)}+\eta_{*}\bm{u}_{*},\label{eq:g0-1}
\intertext{and thus}
\bm{g}^{-}(0) & =\sum_{q=1}^{2N-1}e^{-\lambda_{q}/2}\{\eta_{q}\bm{u}_{(q)}^{-}+\eta_{-q}\bm{u}_{(-q)}^{-}\}+\eta_{0}\bm{u}_{(0)}^{-}+\eta_{*}\bm{u}_{*}^{-}.\nonumber 
\end{align}
Since $\bm{u}_{(q)}=\bm{u}_{(-q)}^{-}$,
the second equation above is rewritten as
\begin{equation}
\bm{g}^{-}(0)=\sum_{q=1}^{2N-1}e^{-\lambda_{q}/2}\{\eta_{q}\bm{u}_{(-q)}+\eta_{-q}\bm{u}_{(q)}\}+\eta_{0}\bm{u}_{(0)}+\eta_{*}\bm{u}_{*}^{-},\label{eq:g-0}
\end{equation}
where $\bm{u}_{(0)}=\bm{u}_{(0)}^{-}$ has been used as well. Since
$\bm{u}_{*}\perp\bm{u}_{(0)}$, it also holds that $\bm{u}_{*}^{-}\perp\bm{u}_{(0)}^{-}$;
thus $\bm{u}_{*}^{-}$ is also perpendicular to $\bm{u}_{(0)}$. Now
let $\bm{u}_{*}^{-}$ expressed as
\[
\bm{u}_{*}^{-}=\zeta_{0}\bm{u}_{(0)}+\zeta_{*}\bm{u}_{*}+\sum_{q=1}^{2N-1}(\zeta_{-q}\bm{u}_{(-q)}+\zeta_{q}\bm{u}_{(q)}).
\]
Since $\{\bm{u}_{(q)}^{-}\}$ and $\{\bm{u}_{(q)}\}$ span the same
vector space, neither $\bm{u}_{*}$ nor $\bm{u}_{*}^{-}$ belongs
to that space. Hence, by putting the first two terms on the right-hand
side to the left-hand side:
\[
\bm{u}_{*}^{-}-\zeta_{*}\bm{u}_{*}-\zeta_{0}\bm{u}_{(0)}=\sum_{q=1}^{2N-1}(\zeta_{-q}\bm{u}_{(-q)}+\zeta_{q}\bm{u}_{(q)}),
\]
it is found that $\zeta_{\pm q}=0$ ($q=1,\dots,2N-1$). Then, the
inner product with $\bm{u}_{(0)}$ shows that $\zeta_{0}=0$. Therefore,
\[
\bm{u}_{*}^{-}=\zeta_{*}\bm{u}_{*},
\]
and the substitution into (\ref{eq:g-0}) gives
\[
\bm{g}^{-}(0)=\sum_{q=1}^{2N-1}e^{-\lambda_{q}/2}\{\eta_{q}\bm{u}_{(-q)}+\eta_{-q}\bm{u}_{(q)}\}+\eta_{0}\bm{u}_{(0)}+\eta_{*}\zeta_{*}\bm{u}_{*}.
\]
Finally, comparing with (\ref{eq:g0-1}) and taking account of the
property $\bm{g}(0)=-\bm{g}^{-}(0)$, the following relations are
obtained:
\[
\eta_{0}=0,\ \eta_{*}=-\eta_{*}\zeta_{*},\ \eta_{q}=-\eta_{-q}\quad(q=1,\dots,2N-1).
\]
Note that $\bm{u}_{*}$ is a unit vector and thus $|\zeta_{*}|=1$.
It is numerically checked that $\bm{u}_{*}^{-}=-\bm{u}_{*}$ actually,
i.e., $\zeta_{*}=-1$. Therefore, $\eta_{*}$ and $\eta_{q}$ ($q=1,\dots2N-1$)
still remains unknown. In order to determine them, finally consider
the condition at $x_{1}=-1/2$. To this end, use the expression
\begin{align*}
\bm{g}(-1/2) & =\sum_{q=1}^{2N-1}\eta_{q}e^{-\lambda_{q}}\bm{u}_{(q)}+\sum_{q=1}^{2N-1}\eta_{-q}\bm{u}_{(-q)}-\frac{1}{2}\eta_{*}\xi\bm{u}_{(0)}+\eta_{*}\bm{u}_{*}\\
 & =\sum_{q=1}^{2N-1}\eta_{q}\{e^{-\lambda_{q}}\bm{u}_{(q)}-\bm{u}_{(q)}^{-}\}-\frac{1}{2}\eta_{*}\xi\bm{u}_{(0)}+\eta_{*}\bm{u}_{*},
\end{align*}
and take its components with a positive subscript,
where $\bm{g}=^{t}[g_{-2N,}\dots,g_{-1},g_{1},\dots,g_{2N}]$.
Then, by the condition at $x_{1}=-1/2$, 
the components $g_1,\dots,g_{2N}$ at $x_{1}=-1/2$ are all unity and thus 
the above expression gives $2N$ equations for $2N$ unknown constants $\eta_{*}$ and $\eta_{q}$ ($q=1,\dots2N-1$). 
Thus the construction of the numerical procedure is completed.

To summarize, in the construction process, it is optimistically supposed
that 
\begin{enumerate}
\item $\mathrm{rank}A=4N$;
\item $\widetilde{D}_{0}\ne0$;
\item $\overline{A}$ is regular;
\item the multiplicity of the eigenvalue zero is two and nonzero eigenvalues
are not multiple: $\lambda_{p}\ne\lambda_{q}$ for $p\ne q$.
\end{enumerate}
These properties have been confirmed numerically to be valid, so that
the constructed procedure has worked well actually.

Before closing this section, there are two things that should be remarked.
Firstly, the present method is applicable for infinite-range potentials
with $\gamma>-2$ only, since the piecewise quadratic approximation of $g$
does not guarantee the continuity of its derivative with respect to $\theta$. Secondly,
on the boundary $x_{1}=-1/2$, the boundary condition is adopted to
represent the value of $g_{+0}$ in the computation, since $g_{+0}$
for $x_{1}=-1/2$ does not necessarily coincides with the value of
$g_{0}\equiv g_{-0}$. It is, however, expected that $g_{\pm0}$ are
the same for $-3<\gamma\le-1$, because of the regularizing effect
of the grazing collision. Indeed, the computed $g_{0}$ is very close
to $g_{+0}$, and furthermore, as the grid intervals are refined,
the tiny difference of the computed $g_{0}$ from $g_{+0}$ tends
to vanish.

\section{Results and Discussions\label{sec:results}}

\subsection{Numerical Results\label{subsec:prerequisit}}

According to the literature, e.g., Refs.~\cite{TT17,TF13,CLT14,TST19,SO73},
 in the case of a hard-sphere gas and the relaxation-type models 
[e.g., the Bhatnagar--Gross--Krook (BGK), the Ellipsoidal Statistical (ES)
model], the velocity distribution function has a jump discontinuity
on the boundary in the molecular velocity space in the direction parallel
to the boundary, which causes the diverging derivative of moment in
the normal direction in approaching the boundary (the moment singularity,
for short). In the case of the flat boundary, the diverging rate
is logarithmic in the distance from the boundary \cite{TF13,CLT14,S64,SO73},
which was first pointed out in the analyses of the Rayleigh problem  by Sone \cite{S64} and of the structure of the Knudsen layer \cite{SO73} 
on the basis of the BGK model. 
The essence of the logarithmic moment singularity
can be understood by the damping model in Ref.~\cite{TF13} that is based
on the strong damping of the jump discontinuity on the boundary by
the loss term for the finite-range potential. 
The jump discontinuity and logarithmic moment singularity
for the finite-range and the cutoff potential 
are the key tests of the present approach via of the Lorentz-gas model. 
\begin{figure}
\centering
\includegraphics[width=0.48\textwidth]{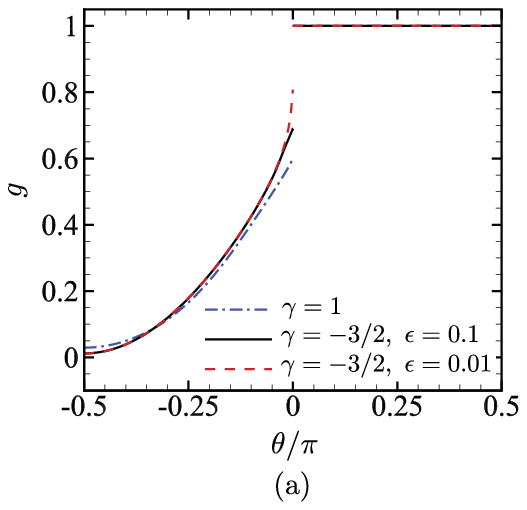}\quad
\includegraphics[width=0.48\textwidth]{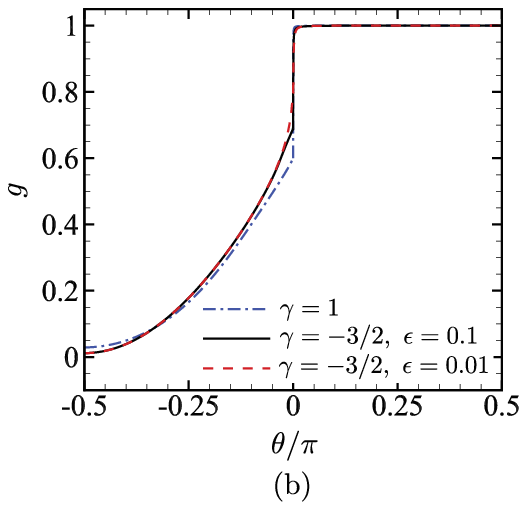}\quad
\includegraphics[width=0.48\textwidth]{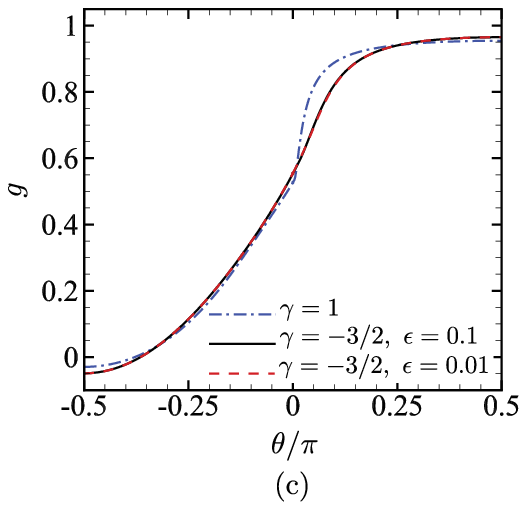}
\caption{Reduced VDF $g$ for the finite-range ($\gamma=1$) and the cutoff potential
($\gamma=-3/2$) with $\epsilon=0.1$ and 0.01. (a) $x_{1}=-1/2$,
(b) $x_{1}=-1/2+0.106\times10^{-3}$, and (c) $x_{1}=-1/2+0.520\times10^{-1}$.\label{fig:g1}}
\end{figure}

\begin{figure}
\centering\includegraphics[width=0.48\textwidth]{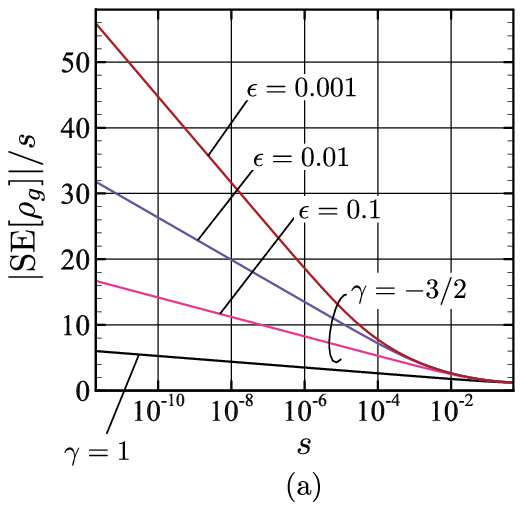}\quad\includegraphics[width=0.48\textwidth]{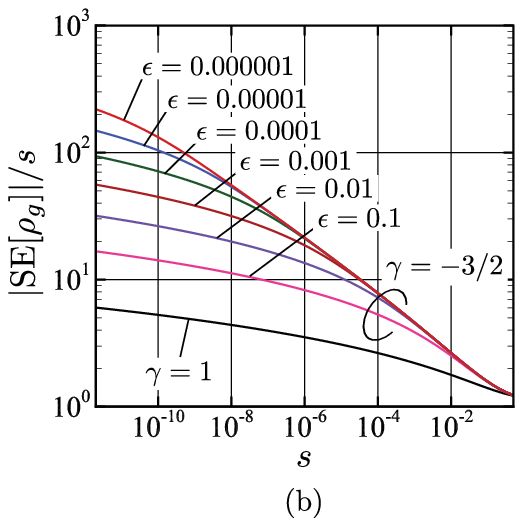}\quad\includegraphics[width=0.48\textwidth]{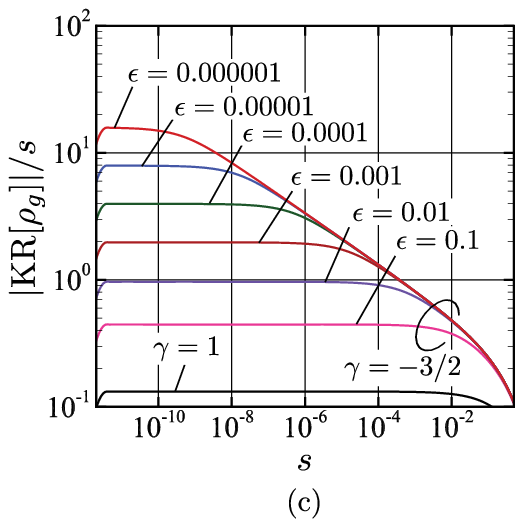}
\caption{Variations of $\rho_{g}$ near the boundary as a function of the normal
distance $s\equiv x_{1}+1/2$ from the boundary for the finite-range
($\gamma=1$) and the cutoff potential ($\gamma=-3/2$) with various sizes of cutoff $\epsilon$. (a) $|\mathrm{SE}[\rho_{g}]|/s$ in the semilog plot, (b) $|\mathrm{SE}[\rho_{g}]|/s$
in the log-log plot, and (c) $|\mathrm{KR}[\rho_{g}]|/s$. \label{fig:g1cuttoff}}
\end{figure}

Figure~\ref{fig:g1} shows the profiles of $g$ for the finite-range potential
with $\gamma=1$ (the hard-disk) and the cutoff potential with $\gamma=-3/2$ (the cutoff Maxwell molecule). As is seen in Fig.~\ref{fig:g1}(a), there is
a jump discontinuity at $\theta=0$ on the boundary $x_{1}=-1/2$,
which vanishes even immediately away from the boundary {[}Figs.~\ref{fig:g1}(b)
and (c){]}. Figure~\ref{fig:g1cuttoff}(a) shows the profile of $\rho_{g}$,
more precisely $|\mathrm{SE}[\rho_{g}]|=|\rho_{g}(x_{1})-\rho_{g}(-1/2)|$
divided by the distance from the boundary (see Appendix~\ref{sec:koike}),
near the boundary for the same case as Fig.~\ref{fig:g1} with the
abscissa being the logarithmic scale. 
Because it shows a nearly straight line for $s\equiv x_{1}+1/2\lesssim10^{-6}$,
$\mathrm{SE}[\rho_{g}]$ (or $\rho_{g}$) changes in proportion
to $s\ln s$ from its value on the boundary. In other words, $d\rho_{g}/dx_{1}$
diverges logarithmically in approaching the boundary. Hence, the moment
singularity studied in Refs.~\cite{TF13,CLT14,SO73,CH15} is well reproduced
by the present Lorentz-gas model. 

\begin{figure}
\centering\includegraphics[width=0.48\textwidth]{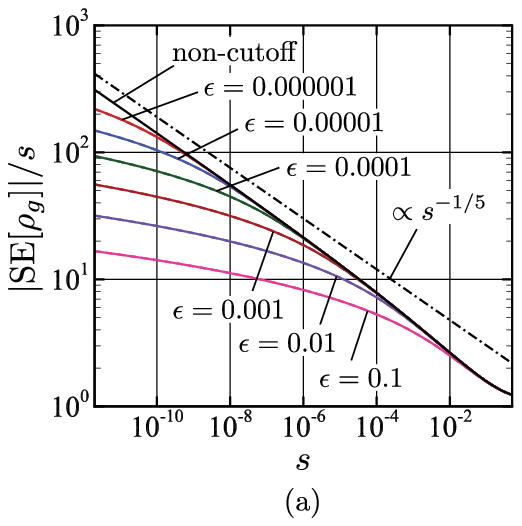}\quad\includegraphics[width=0.48\textwidth]{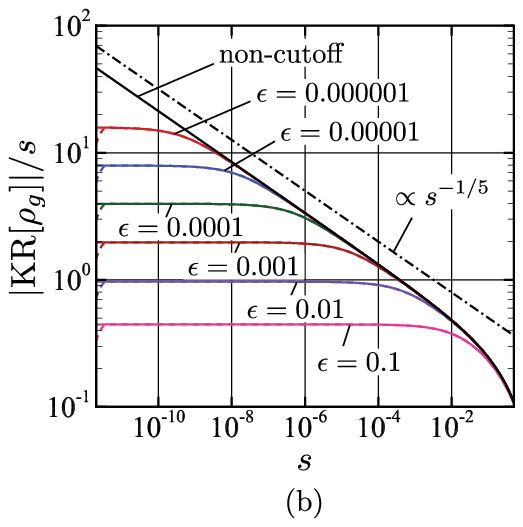}
\caption{Variations of $\rho_{g}$ near the boundary as a function of the normal
distance $s\equiv x_{1}+1/2$ from the boundary for the infinite-range
and the corresponding cutoff potential ($\gamma=-3/2$). 
(a) $|\mathrm{SE}[\rho_{g}]|/s$ and (b) $|\mathrm{KR}[\rho_{g}]|/s$.
The solid lines indicate the results by the Galerkin method. 
The dashed lines the results by the direct method. 
The latter agree well with the former and are almost invisible
except for the left end in (b).\label{fig:g-3/2noncuttoff}}
\end{figure}

Next, the results for the cutoff potential with $\gamma=-3/2$ for
various values of $\epsilon$ down to $10^{-6}$ from $10^{-1}$ are
shown in Figs.~\ref{fig:g1cuttoff}(b) and \ref{fig:g1cuttoff}(c). Again, $|\mathrm{SE}[\rho_{g}]|$
divided by the distance from the boundary is shown in Fig.~\ref{fig:g1cuttoff}(b),
but as the log-log plot. It is observed that the profiles for different
$\epsilon$ forms an envelope outside the region of logarithmic change
in Fig.~\ref{fig:g1cuttoff}(a) and that the envelope extends towards
the boundary as $\epsilon$ decreases. Although it is not enough clear
in Fig.~\ref{fig:g1cuttoff}(b), the envelope follows the power law
of the distance $s$, which is clearly demonstrated in Fig.~\ref{fig:g1cuttoff}(c),
where $|\mathrm{KR}[\rho_{g}]|$ (in place of $|\mathrm{SE}[\rho_{g}]|$)
divided by the distance is shown as the log-log plot, following an efficient 
estimate method by Koike \cite{K21} (see Appendix~\ref{sec:koike} for the
definition of $\mathrm{KR}$), in order to pick up the asymptotic
behavior of $\rho_{g}$ near the boundary efficiently. The envelope
part becomes nearly straight in Fig.~\ref{fig:g1cuttoff}(c) with
its slope very close to $-1/5$;\footnote{The horizontal straight part shows that $|\mathrm{KR}[\rho_{g}]|$
divided by the distance $s$ is proportional to $\ln s$ there.} $|\mathrm{KR}[\rho_{g}]|$ divided by the distance is proportional
to $s^{-1/5}$ there. Furthermore, the envelope extends again toward
the boundary as $\epsilon\to0$. This strongly suggests that, for
the infinite range potential, the logarithmic divergence observed
in the cutoff potential does not occur and instead the diverging
rate becomes stronger, here $s^{-1/5}$ for $\gamma=-3/2$. In order
to confirm it, the computation for the infinite-range potential with
$\gamma=-3/2$ has been carried out by the Galerkin method. The result
is shown in Fig.~\ref{fig:g-3/2noncuttoff}. The results obtained
by the Galerkin method applied to the cutoff potential are also shown
for comparisons with those obtained by the direct method for the reliability
assessment of both methods. Excellent agreement is achieved both in
Figs.~\ref{fig:g-3/2noncuttoff}(a) and \ref{fig:g-3/2noncuttoff}(b). As expected, the envelope
extends indeed down to the boundary for the infinite-range potential.
From Fig.~\ref{fig:g-3/2noncuttoff}(b), the slope of $|\mathrm{KR}[\rho_{g}]|$
divided by the distance is estimated as $-1/5$. This confirms that
$d\rho_{g}/dx_{1}$ diverges with the rate $s^{-1/5}$ in approaching
the boundary (i.e., as $s\to0$). 

Incidentally, the computation of $|\mathrm{KR}[\rho_{g}]|$ can be
sensitive to the round off errors, compared with the simpler computation
of $|\mathrm{SE}[\rho_{g}]|$. Accordingly, the unnatural change of
profile is observed for very small value of $s$ in the results of
the direct method, because its numerical code makes use of the double
precision arithmetic. Such unnatural behavior is not observed in the
results of the Galerkin method, where the numerical code fully makes
use of the multiple precision arithmetic with the aid of efficient
libraries: exflib \cite{F} by Fujiwara and Python-FLINT \cite{J} by
Johansson.

\subsection{Discussions\label{subsec:discussion}}

\begin{figure}
\centering
\includegraphics[width=0.48\textwidth]{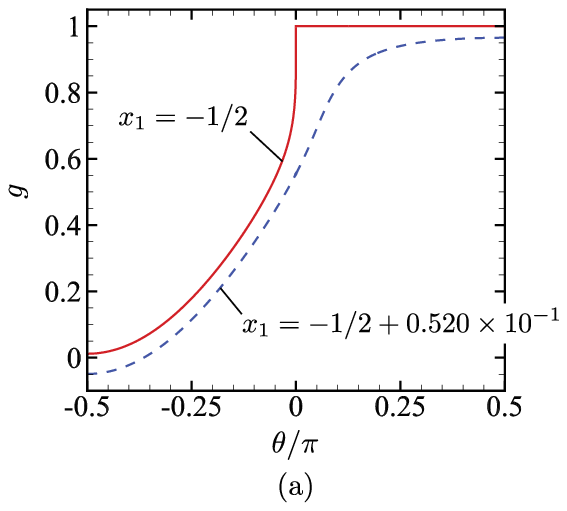}
\quad
\includegraphics[width=0.48\textwidth]{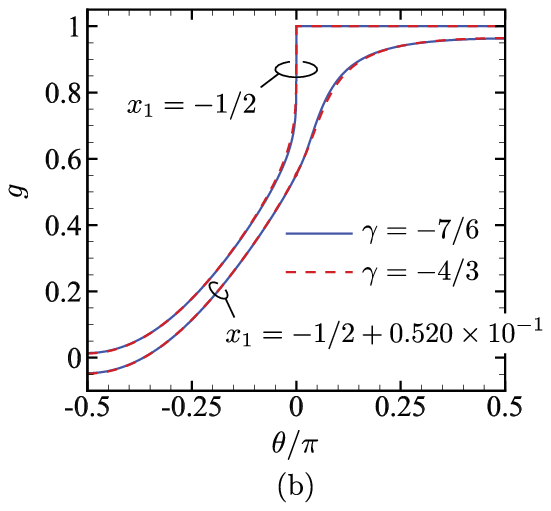}
\caption{Reduced VDF $g$ for the infinite-range potential on and away from the boundary.
(a) $\gamma=-3/2$, (b) $\gamma=-4/3$ and $-7/6$. \label{fig:g1-1}}
\end{figure}

In viewing the existing works for the finite-range potential, the
diverging gradient of macroscopic quantities originates from
the jump discontinuity of the VDF on the boundary. In this sense, it is
striking that the singularity of diverging gradient occurs (more strongly) for the infinite-range
potential in spite of the fact that the grazing collision regularizes $g$
to have no jump discontinuity on the boundary as shown in Fig.~\ref{fig:g1-1}(a);
see also Fig.~\ref{fig:g1-1}(b) for other values of $\gamma$.
We show below two clue observations that give the hints to this unexpected
result.
\begin{figure}
\centering\includegraphics[width=0.48\textwidth]{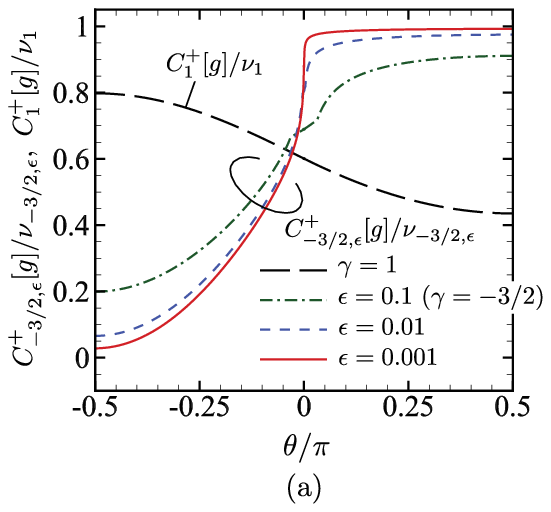}\quad
\includegraphics[width=0.48\textwidth]{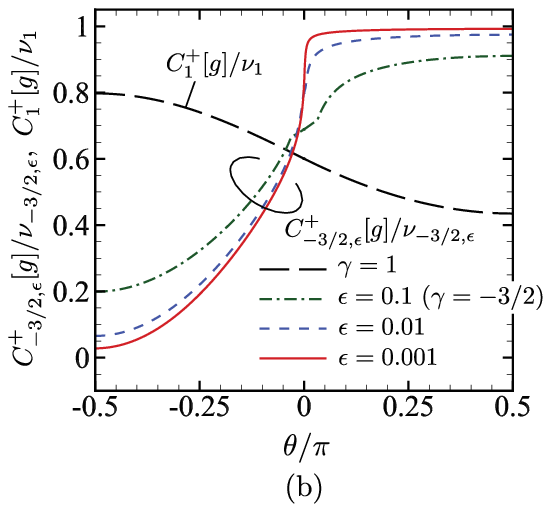}\quad
\includegraphics[width=0.48\textwidth]{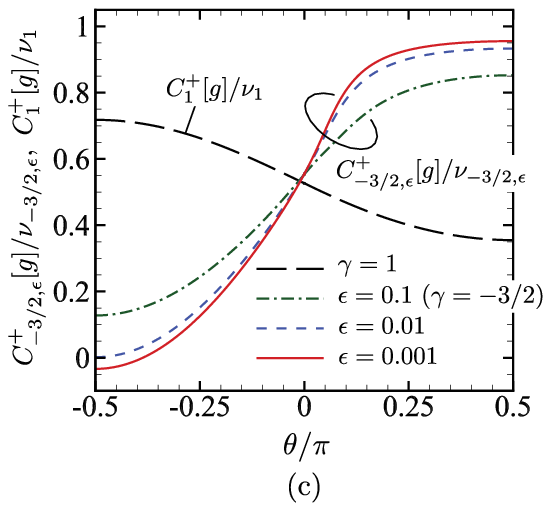}
\caption{Gain term divided by the collision frequency for the finite-range ($\gamma=1$) and the cutoff potential ($\gamma=-3/2$) with $\epsilon=0.1, 0.01,0.001$: $C_{-3/2,\epsilon}^{+}[g]/\nu_{-3/2,\epsilon}$ and $C_{1}^{+}[g]/\nu_{1}$.
(a) $x_{1}=-1/2$, (b) $x_{1}=-1/2+0.106\times10^{-3}$, and (c) $x_{1}=-1/2+0.520\times10^{-1}$.
\label{fig:fig3}}
\end{figure}

\begin{figure}
\centering\includegraphics[width=0.48\textwidth]{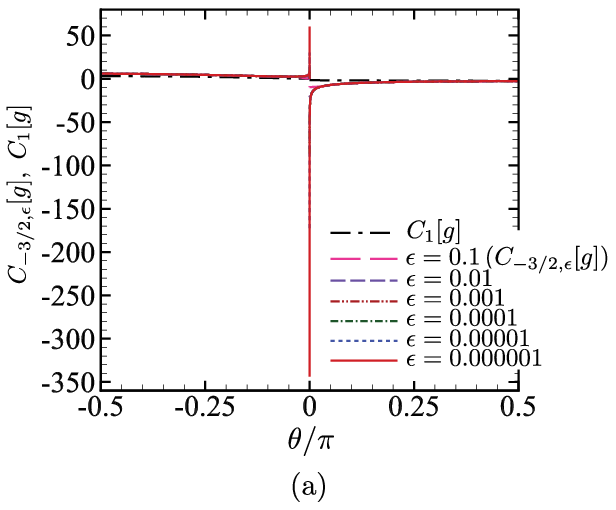}\quad
\includegraphics[width=0.48\textwidth]{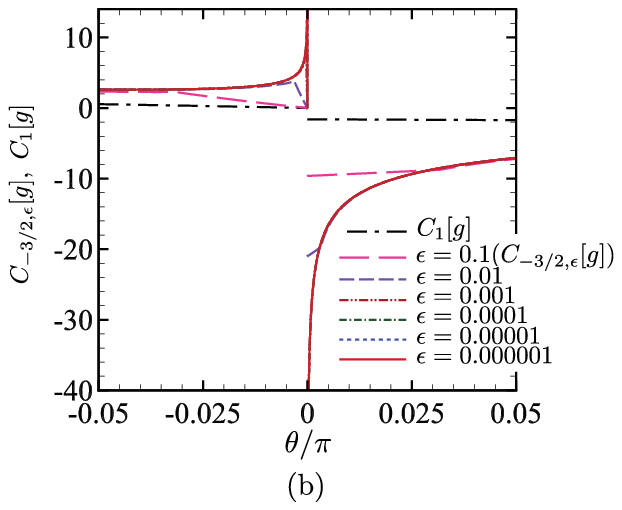}\quad
\includegraphics[width=0.48\textwidth]{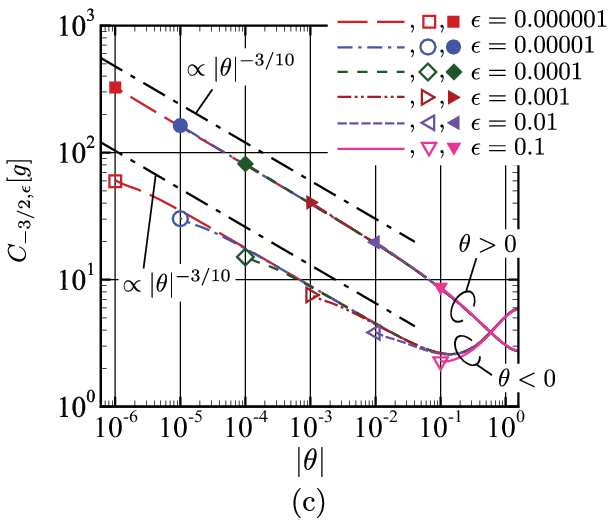}
\caption{Collision integral for the finite-range ($\gamma=1$) and the cutoff potential ($\gamma=-3/2$) on the boundary $x_1=-1/2$: $C_{-3/2,\epsilon}[g]$ and $C_{1}[g]$. 
(a) $C_{-3/2,\epsilon}[g]$  and $C_{1}[g]$ as functions of $\theta/\pi$,
(b) close-up of (a), and (c) $C_{-3/2,\epsilon}[g]$ for $|\theta|\ge\epsilon$. 
In (c), the data  $C_{-3/2,\epsilon}[g]$ are plotted for various values of $\epsilon$
and the data at $|\theta|=\epsilon$ are indicated by a symbol.
\label{fig:fig4}}
\end{figure}

The first clue is the collision term $C[g]$. For the finite-range
potential, the singular feature of $C[g]$ is confined in the loss
term as the jump discontinuity of $g$ and the gain term $C^{+}[g]$
behaves smoothly as demonstrated in Fig.~\ref{fig:fig3} (see the
case $\gamma=1$). For the cutoff potential, however, $C^{+}[g]$
changes steeply for $\theta\sim0$, losing the smooth feature observed
for the finite range potential (see Fig.~\ref{fig:fig3} for $\gamma=-3/2$
with small $\epsilon$). Accordingly, even after combined with the
loss term, the collision integral $C[g]$ changes steeply
and tends to diverge as $\theta\to0$; see Figs.~\ref{fig:fig4}(a)
and \ref{fig:fig4}(b). 
Figure~\ref{fig:fig4}(c) shows the behavior of $C_{-3/2,\epsilon}[g]$ 
on the boundary for various values of $\epsilon$,
which strongly suggests that $C_{-3/2}[g]$ on the boundary diverges
in the limit $\theta\to0$ with the rate $|\theta|^{-3/10}$.\footnote{The diverging rate is expected to be $|\theta|^{\gamma\frac{\gamma+1}{\gamma-1}}$ (or $|\theta|^{\gamma/n}$) by additional observations for other values of $\gamma$ in $]-3,-1[$, though they are omitted in the present paper.} 
The grazing collision induces, even if locally, the divergence of the collision integral, as the price for regularizing the VDF. 
The trade-off makes the situation worse in the moment singularity.

\begin{figure}
\centering\includegraphics[width=0.48\textwidth]{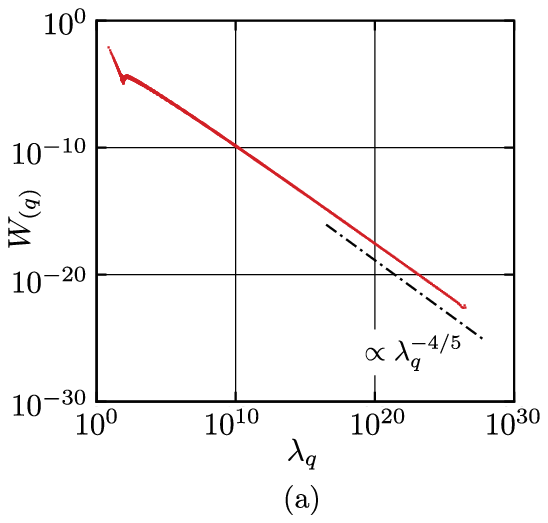}\quad\includegraphics[width=0.48\textwidth]{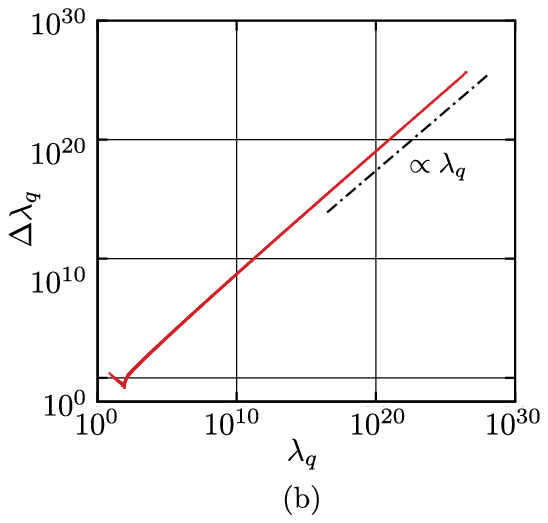}

\caption{Weight $W_{(q)}$ and the interval of eigenvalues $\Delta\lambda_{q}$
against the eigenvalue $\lambda_{q}$ for the infinite-range potential ($\gamma=-3/2$).
(a) $W_{(q)}$ and (b) $\Delta\lambda_{q}$.
In the plot, the data for the grid system for $\theta$ with $N=688$ are used,
where the grid points next to the origin are $\theta^{(\pm1)}=\pm 1.05\times10^{-17}$. 
\label{fig:fig5}}
\end{figure}

The second clue is the correspondence among the eigenvalues $\lambda_{\pm1},\cdots,\lambda_{\pm(2N-1)}$
and coefficients $\eta_{\pm1},\cdots,\eta_{\pm(2N-1)}$ that occur
in the exponential elements; see (\ref{eq:g_dc}). Thanks to (\ref{eq:basisFunc}),
$\rho_{g}$ is expressed as
\[
\rho_{g}(x_{1})=\sum_{j=-2N}^{2N}g_{j}(x_{1})w_{j},\quad w_{j}=\frac{1}{\pi}\int_{-\pi/2}^{\pi/2}Y_{j}(\theta)d\theta.
\]
Then, the substitution of \eqref{eq:g0} gives
 \[
 \rho_{g}(x_{1})
=\sum_{j\ne0}g_{j}(x_{1})\{w_{j}-\frac{\widetilde{D}_j}{\widetilde{D}_0}w_0\},
\]
which is further transformed by the substitution of (\ref{eq:g_dc}) as follows:%
\begin{align*}
\rho_{g}(x_{1}) & =\sum_{j\ne0}\Big[\sum_{q=1}^{2N-1}\{\eta_{q}e^{\lambda_{q}(x_{1}-\frac{1}{2})}u_{(q)j}+\eta_{-q}e^{-\lambda_{q}(x_{1}+\frac{1}{2})}u_{(-q)j}\}\displaybreak[0] \\
&\qquad\qquad +\eta_{*}\xi x_{1}u_{(0)j}+\eta_{*}u_{*j}\Big]\{w_{j}-\frac{\widetilde{D}_j}{\widetilde{D}_0}w_0\}\displaybreak[0]\\
 & =\sum_{q=1}^{2N-1}\{W_{(q)}e^{\lambda_{q}(x_{1}-\frac{1}{2})}+W_{(-q)}e^{-\lambda_{q}(x_{1}+\frac{1}{2})}\}+\xi x_{1}W_{(0)}+W_{*}\displaybreak[0]\\
 & =\sum_{q=1}^{2N-1}W_{(q)}\{e^{\lambda_{q}(x_{1}-\frac{1}{2})}-e^{-\lambda_{q}(x_{1}+\frac{1}{2})}\}+\xi x_{1}W_{(0)}+W_{*},
\intertext{where}%\displaybreak[0]
 & W_{(q)}=\sum_{j\ne0}\{w_{j}-\frac{\widetilde{D}_j}{\widetilde{D}_0}w_0\}\eta_{q}u_{(q)j}=-W_{(-q)},\displaybreak[0]\\
 & W_{(0)}=\sum_{j\ne0}\{w_{j}-\frac{\widetilde{D}_j}{\widetilde{D}_0}w_0\}\eta_{*}u_{(0)j},\quad  W_{*}=\sum_{j\ne0}\{w_{j}-\frac{\widetilde{D}_j}{\widetilde{D}_0}w_0\}\eta_{*}u_{*j},
\end{align*}%
and  $\eta_{0}=0$ has been used. Figure~\ref{fig:fig5} shows $W_{(q)}$
vs $\lambda_{q}$ and $\Delta\lambda_{q}$ vs $\lambda_{q}$ for the
infinite-range potential with $\gamma=-3/2$, where $\Delta\lambda_{q}=\lambda_{q}-\lambda_{q-1}$
and $\lambda_{q}$ increases indefinitely as $q\to\infty$. From the
figure, it is seen that $W_{(q)}\propto\lambda_{q}^{-4/5}$ and $\Delta\lambda_{q}\propto\lambda_{q}$
as $\lambda_{q}$ (or $q$) increases. Then, as is often done in the
statistical mechanics for large $N$, the summation with respect to $q$ is well estimated by the integration as $\sum_{q=1}^{2N-1}W_{(q)}e^{-a\lambda_{q}}=\int_{\lambda_1}^{\infty}W(\lambda)e^{-a\lambda}d\lambda$
for $a>0$, where $W$, $\lambda$, and $d\lambda$ are the appropriate
continuous counterparts of $W_{(q)}/\Delta\lambda_{q}$, $\lambda_{q}$,
and $\Delta\lambda_{q}$. For the present purpose
of the diverging rate estimate, the lower bound of the integration range $\lambda_1$ may be replaced by unity, because only the behavior of the integrand for large $\lambda$ is relevant. 

Hence, because of Fig.~\ref{fig:fig5}, $W(\lambda)\sim\lambda^{-9/5}$ for $\gamma=-3/2$,
and the singular behavior of $\rho_{g}$ can be estimated by
\[
\int_{1}^{\infty}\lambda^{-9/5}\exp(-\lambda s)d\lambda=\frac{5}{4}-\frac{5\pi\sec(\frac{3\pi}{10})}{4\Gamma(\frac{4}{5})}s^{4/5}+O(s).
\]
By taking the derivative with respect to $s$, the diverging rate
$s^{-1/5}$ of $d\rho_{g}/dx_{1}$ is reproduced.

\subsection{Conjecture on the Diverging Rate for Infinite-Range Potentials\label{subsec:Conjecture}}

\begin{figure}
\centering\includegraphics[width=0.48\textwidth]{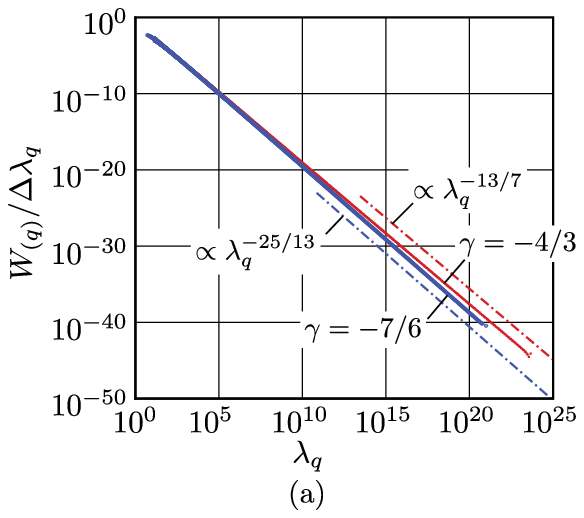}\quad
\includegraphics[width=0.48\textwidth]{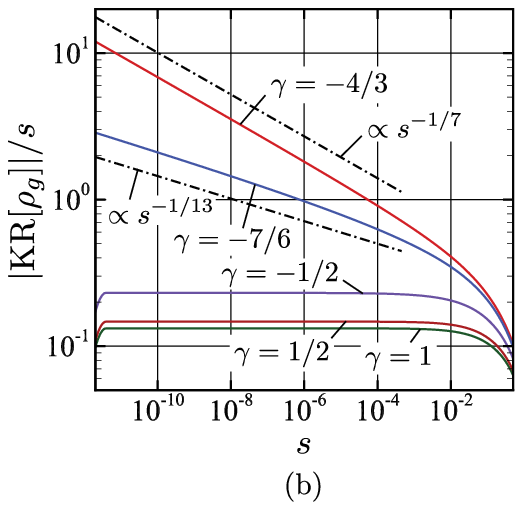}
\caption{$W_{(q)}/\Delta\lambda_{q}$
against the eigenvalue $\lambda_{q}$
and the variation of $\rho_g$ near the boundary for the infinite-range potential in the case $\gamma=-4/3,-7/6$.
(a) $W_{(q)}/\Delta\lambda_{q}$ and (b) $|\mathrm{KR}[\rho_g]|$.
In (b) the cases for the finite-range potential ($\gamma=1$ and $\gamma=\pm1/2$) are also shown for reference.
For (a), see the caption of Fig.~\ref{fig:fig5} as well.
\label{fig:fig6}}
\end{figure}

From the detailed observations on the case $\gamma=-3/2$, it is conjectured
for $\gamma<-1$ that

\begin{equation}
W(\lambda)\sim\lambda^{\frac{2}{\gamma-1}-1}=\lambda^{\frac{1}{n}-2}\quad\mbox{as }\lambda\to\infty,\label{eq:W_conj}
\end{equation}
and that the diverging rate of $d\rho_g/dx_1$ is $s^{-\frac{\gamma+1}{\gamma-1}}=s^{-1/n}$.
Indeed, this conjecture recovers the second clue part of Sec.~\ref{subsec:discussion}. %When $\gamma=-1$ (or $n=\infty$), 
%(\ref{eq:W_conj}) gives
%\[
%W(\lambda)\sim\lambda^{-2}\quad\mbox{as }\lambda\to\infty,
%\]
%\[
%\int_{1}^{\infty}\lambda^{-2}\exp(-\lambda s)d\lambda\sim1+s\ln s+O(s),
%\]
%and predicts the diverging rate of $\ln s$; w
When $\gamma=-7/6$
(or $n=13)$, it gives
\[
W(\lambda)\sim\lambda^{-25/13}\quad\mbox{as }\lambda\to\infty,
\]
\[
\int_{1}^{\infty}\lambda^{-25/13}\exp(-\lambda s)d\lambda=\frac{13}{12}-\frac{13\pi\sec(\frac{11}{26}\pi)}{12\Gamma(\frac{12}{13})}s^{12/13}+O(s),
\]
and predicts the diverging rate of $s^{-1/13}$; when $\gamma=-4/3$ (or
$n=7)$, it gives
\[
W(\lambda)\sim\lambda^{-13/7}\quad\mbox{as }\lambda\to\infty,
\]
\[
\int_{1}^{\infty}\lambda^{-13/7}\exp(-\lambda s)d\lambda=\frac{7}{6}-\frac{7\pi\sec(\frac{5}{14}\pi)}{6\Gamma(\frac{6}{7})}s^{6/7}+O(s),
\]
and predicts the diverging rate of $s^{-1/7}$. The prediction
rates for $\gamma=-4/3,-7/6$ are also confirmed numerically, as shown
in Fig.~\ref{fig:fig6}. 

\begin{figure}
\centering\includegraphics[width=0.48\textwidth]{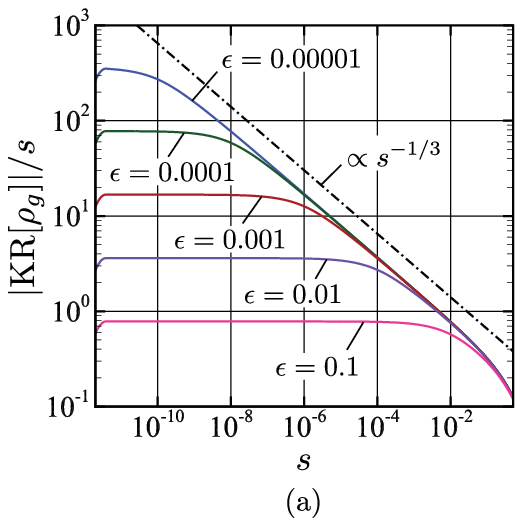}\quad
\includegraphics[width=0.48\textwidth]{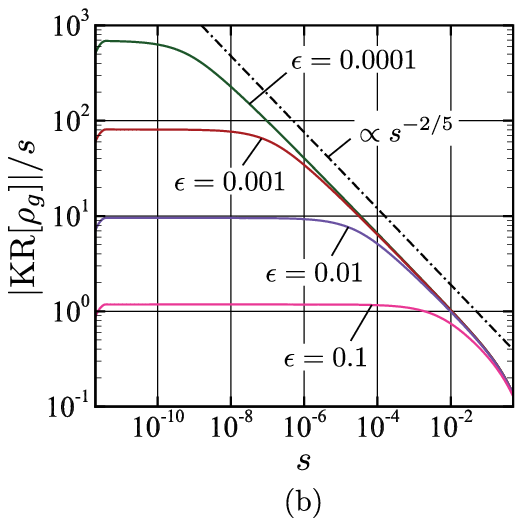}
\caption{Variation of $\rho_g$ for the cutoff potential for $\gamma=-2$ and $-7/3$ with various sizes of cutoff $\epsilon$. 
(a) $\gamma=-2$ and (b) $\gamma=-7/3$.
See \eqref{eq:KR} in Appendix~\ref{sec:koike} for $\mathrm{KR}[\rho_g]$.
\label{fig:fig9}}
\end{figure}

Furthermore, when $\gamma=-2$ (or $n=3$),
it gives
\[
W(\lambda)\sim\lambda^{-5/3}\quad\mbox{as }\lambda\to\infty,
\]
\[
\int_{1}^{\infty}\lambda^{-5/3}\exp(-\lambda s)d\lambda
=\frac{3}{2}-\frac{\sqrt{3}\pi}{\Gamma(\frac{2}{3})}s^{2/3}+O(s),
\]
and predicts the diverging rate of $s^{-1/3}$; when $\gamma=-7/3$ (or $n=5/2$),
it gives
\[
W(\lambda)\sim\lambda^{-8/5}\quad\mbox{as }\lambda\to\infty,
\]
\[
\int_{1}^{\infty}\lambda^{-8/5}\exp(-\lambda s)d\lambda
=\frac{5}{3}-\frac{5\pi\sec(\frac{\pi}{10})}{3\Gamma(\frac{3}{5})}s^{3/5}+O(s),
\]
and predicts the diverging rate of $s^{-2/5}$.
Although the direct numerical assessment is not available 
for $\gamma\le-2$ at present,
an alternative assessment is possible 
by numerically observing the asymptotic behavior of the \textit{envelope}
in $|\mathrm{KR}[\rho_g]|/s$ for small $\epsilon$'s by using the direct method;
the results support the prediction for $\gamma=-2$ and $-7/3$;
see Fig.~\ref{fig:fig9}.

 
To summarize, the diverging rate is logarithmic for the finite-range ($-1<\gamma\le1$)
and the cutoff potential 
[see Fig.~\ref{fig:g1cuttoff} and Fig.~\ref{fig:fig6}(b) for $\gamma=\pm1/2$], while it is $s^{-\frac{\gamma+1}{\gamma-1}}=s^{-1/n}$ 
for the infinite-range potential with $-3<\gamma<-1$.%
\footnote{For $\gamma=-1$, the above conjecture predicts the logarithmic rate. This setting is, however, not realized by a fixed value of $n$, but realized only in the limit $n\to\infty$. The case $\gamma=-1$ is thus marginal. Indeed, the decisive evidence was not obtained numerically by the direct method for the cutoff case, 
even from the data ranging from $\epsilon=10^{-1}$ down to $10^{-9}$.}

\section{Conclusion\label{sec:conclusion}}

Using a mono-speed Lorentz-gas model, the moment singularity near the \textit{flat} boundary has been investigated. 
First, the logarithmic moment
singularity in approaching the boundary is checked to be reproduced
for the finite-range and the cutoff potentials by the Lorentz-gas
model. The jump discontinuity of the velocity distribution function
is also reproduced well on the boundary. Then, by using
the Galerkin method for the infinite-range potential, it is demonstrated
that the grazing collision indeed has the regularizing effect on the
velocity distribution function and that the jump discontinuity disappears
on the boundary. Surprisingly however, the moment singularity is
not weakened but rather strengthened to be of the inverse power of
the distance from the boundary. This is due to the fact that the collision
integral becomes locally infinite in the molecular velocity direction
parallel to the boundary ($\theta=0$) as the price for the regularization
of the VDF on the boundary. By detailed analyses of the high-resolution
numerical data, a conjecture is made for the prediction of
the diverging rate for the infinite range potential with $-3<\gamma<-1$,
which are numerically confirmed for different values of $\gamma$.
In conclusion, the diverging rate is logarithmic for the finite-range ($-1<\gamma\le1$)
and the cutoff potential, while it is $s^{-\frac{\gamma+1}{\gamma-1}}=s^{-1/n}$ 
for the infinite-range potential with $-3<\gamma<-1$.

Finally, by the present work, it is strongly suggested
that for the infinite-range potential the collision integral of the
standard Boltzmann equation does not remain finite on the boundary
and that the moment singularity is induced as well near the boundary. 
The rate expected near the planar boundary is of the inverse-power
which is stronger than the logarithmic rate for the finite-range and
the cutoff potential.

\appendix

\section{Basis Functions\label{sec:kernel}}

For the sake of the numerical convenience, the grid points in $\theta$-space
are arranged to be symmetric with respect to $\theta=0$ in the region
$-\pi/2\le\theta\le\pi/2$ so as to make $2N$ small intervals in
both the positive and negative side:
\begin{align*}
0 & =\theta^{(0)}<\theta^{(1)}<\cdots<\theta^{(2N-1)}<\theta^{(2N)}=\pi/2,\quad\theta^{(-j)}=-\theta^{(j)},\quad(j=1,\dots,2N).
\end{align*}
The size of the intervals is not uniform and is smaller near $\theta=0$
so that many grid points are around there. Then the following basis
function set $\{Y_{i}(\theta)\}$ ($i=-2N,\dots,2N$) is used for
the piecewise quadratic approximation of a function of $\theta$:
\[
Y_{2\ell}(\theta)=\begin{cases}
{\displaystyle \frac{(\theta-\theta^{(2\ell+1)})(\theta-\theta^{(2\ell+2)})}{(\theta^{(2\ell)}-\theta^{(2\ell+1)})(\theta^{(2\ell)}-\theta^{(2\ell+2)})}}, & \theta^{(2\ell)}<\theta<\theta^{(2\ell+2)},\ -N\le\ell<N,\\
{\displaystyle \frac{(\theta-\theta^{(2\ell-1)})(\theta-\theta^{(2\ell-2)})}{(\theta^{(2\ell)}-\theta^{(2\ell-1)})(\theta^{(2\ell)}-\theta^{(2\ell-2)})}}, & \theta^{(2\ell-2)}<\theta<\theta^{(2\ell)},\ -N<\ell\le N,\\
0, & \mbox{otherwise},
\end{cases}
\]
\[
Y_{2\ell+1}(\theta)=\begin{cases}
{\displaystyle \frac{(\theta-\theta^{(2\ell)})(\theta-\theta^{(2\ell+2)})}{(\theta^{(2\ell+1)}-\theta^{(2\ell)})(\theta^{(2\ell+1)}-\theta^{(2\ell+2)})}}, & \theta^{(2\ell)}<\theta<\theta^{(2\ell+2)},\ -N\le\ell<N,\\
0, & \mbox{otherwise}.
\end{cases}
\]
By definition, $Y_{j}(\theta)=Y_{-j}(-\theta)$ and that $Y_{0}(\theta)$
is even in $\theta$.

In the direct method, $Y_{\pm0}(\theta)=Y_{0}(\theta)H(\pm\theta)$ is
also prepared to express the jump discontinuity of $g$ at $\theta=0$,
where $H(\theta)$ is the Heaviside function. Using the notation $g_{\pm0}(x_{1})=g(x_{1},\theta=\pm0)$,
the $g$ having a jump discontinuity at $\theta=0$ is approximated
by%
\begin{equation}
g(x_{1},\theta)=\sum_{i=1}^{2N}\{g_{-i}(x_{1})Y_{-i}(\theta)+g_{i}(x_{1})Y_{i}(\theta)\}+g_{-0}(x_{1})Y_{-0}(\theta)+g_{+0}(x_{1})Y_{+0}(\theta).\label{eq:disc_g}
\end{equation}%
If there is no jump discontinuity, $g$ is simply approximated by
$g=\sum_{i=-2N}^{2N}g_{i}Y_{i}(\theta)$ with the simplified notation
$g_{0}(x_{1})\equiv g_{\pm0}(x_{1})$. Accordingly, the numerical
kernel used in the direct method takes the form $C^{+}[g]=\sum_{i=1}^{2N}\{g_{-i}C^{+}[Y_{-i}]+g_{i}C^{+}[Y_{i}]\}+g_{-0}C^{+}[Y_{-0}]+g_{+0}C^{+}[Y_{+0}]$
or $C^{+}[g]=\sum_{i=-2N}^{2N}g_{i}C^{+}[Y_{i}]$, depending on whether
the jump discontinuity exists or not.

The %substitution of $g=\sum_{i=-2N}^{2N}g_{i}Y_{i}(\theta)$
%gives $C^{+}[g]=\sum_{i=-2N}^{2N}g_{i}C^{+}[Y_{i}]$, where the 
analytical expression of $C^{+}[Y_{i}]$ is available with the aid of the series
expansion of $|\sin\frac{\varphi}{2}|^{\gamma}$. 
Although it is truncated by a finite number of terms, 
the expression is helpful to perform the accurate numerical computation. 
The same applies to the Galerkin method, i.e.,
both $A_{ij}$ and $D_{ij}$ can be obtained
analytically as well even for the infinite-range potential.
The highly
accurate computations with the multiple precision arithmetic are achieved
in this way.

\section{Acceleration Method for Estimating the Asymptotic Behavior\label{sec:koike}}

In the present study, an acceleration method proposed in Ref.~\cite{K21}
that makes use of the Richardson extrapolation is found to be very
powerful in estimating the asymptotic behavior of the density in approaching
the boundary. The method is briefly explained in this appendix.

Suppose that a function $f$ of $x(\ge X)$ behaves 
\begin{equation}
f(x)\sim f(X)+a_{\alpha}s^{\alpha}+a_{1}s+o(s),\label{eq:fasymp}
\end{equation}
for $x\sim X$, where $s=x-X$ and $0<\alpha<1$ is an unknown constant.
In the application to the present work, put $X=-1/2$. The idea of
the method is composed of killing the third term to clearly pick up
the second term on the right-hand side, thereby improving the estimate
of the exponent $\alpha$ by the linear regression on the log-log plot.

The straightforward estimate (SE) for the exponent $\alpha$
is just to take
\begin{equation}
\mathrm{SE}[f]\equiv f(x)-f(X)\sim a_{\alpha}s^{\alpha}+a_{1}s+o(s),\label{eq:SE}
\end{equation}
and to use the linear regression. As is clear from the most-right-hand
side, however, the $O(s)$ term may affect the linear regression unless
a clear difference of scale appears in the data at hands. In Ref.~\cite{K21},
the following combination of $f$ that makes use of the Richardson extrapolation is proposed by Koike (the KR method, for short):

\begin{equation}
\mathrm{KR}[f]\equiv f(x)-2f(X+s/2)+f(X).\label{eq:KR}
\end{equation}
Then, it behaves 
\[
\mathrm{KR}[f]\sim a_{\alpha}(1-2^{1-\alpha})s^{\alpha}+o(s),
\]
and accordingly there is no longer influence of the term $O(s)$ in
the linear regression. Hence, the estimate of $\alpha$
should be improved.

Practically, there is a possible drawback such
that $\mathrm{KR}[f]$ would require more significant digits than
$\mathrm{SE}[f]$ in order to avoid the influence of the round-off
error. Indeed, in Figs.~\ref{fig:g1cuttoff}(c) and \ref{fig:g-3/2noncuttoff}(b),
the influence can be observed in the results by the direct method
but not in the results by the Galerkin method. The difference comes
from that the computation code for the former uses the double precision
arithmetic, while that for the latter uses the multiple precision
arithmetic and does not make a discretization in $x_{1}$.

%\section*{Acknowledgements}
\acknowledgement
The present work has been supported in part by the research donation to S.T. 
from Osaka Vacuum Ltd. and by the Japan-France Integrated Action Program
(SAKURA) (Grant No. JSPSBP120193219).
The authors thank Kai Koike for informing them his idea 
of the efficient estimate method \cite{K21}.


\begin{thebibliography}{10}
\bibitem{K69}Kogan, M. N.: Rarefied Gas Dynamics, Plenum Press, New
York (1969).

\bibitem{S07}Sone, Y.: Molecular Gas Dynamics, Birkh\"{a}user, Boston
(2007); supplementary notes and errata are available from KURENAI (http://hdl.handle.net/2433/66098).

\bibitem{TT17}Takata, S., Taguchi, S.: Gradient divergence of fluid-dynamic
quantities in rarefied gases on smooth boundaries, J. Stat. Phys.
\textbf{168}, 1319--1352 (2017). https://doi.org/10.1007/s10955-017-1850-7.

\bibitem{TF13}Takata, S., Funagane, H.: Singular behaviour of a rarefied
gas on a planar boundary, J. Fluid Mech. \textbf{717}, 30--47 (2013).
https://doi.org/10.1017/jfm.2012.559.

\bibitem{CLT14}Chen, I.-K., Liu, T.-P., Takata, S.: Boundary singularity
for thermal transpiration problem of the linearized Boltzmann equation,
Arch. Rational Mech. Anal. \textbf{212}, 575--595 (2014). https://doi.org/10.1007/s00205-013-0714-9.

\bibitem{TST19}Taguchi, S., Saito, K., Takata, S.: A rarefied gas
flow around a rotating sphere: diverging profiles of gradients of
macroscopic quantities, J. Fluid Mech. \textbf{862}, 5--33 (2019).
https://doi.org/10.1017/jfm.2018.946.

\bibitem{CH15}Chen, I.-K., Hsia, C.-H.: Singularity of macroscopic
variables near boundary for gases with cutoff hard potential, SIAM
J. Math. Anal. \textbf{47}, 4332--4349 (2015). 

\bibitem{C88}Cercignani, C.: The Boltzmann Equation and Its Applications,
Springer, Berlin (1988). http://dx.doi.org/10.1007/978-1-4612-1039-9.

\bibitem{D95}Desvillettes, L.: About the regularizing properties
of the non-cut-off Kac equation, Commun. Math. Phys. \textbf{168},
417--440 (1995). https://doi.org/10.1007/BF02101556.

\bibitem{DG00}Desvillettes, L., Golse, F.: On a model Boltzmann equation
without angular cutoff, Differential and Integral Equations \textbf{13},
567--594 (2000).

\bibitem{V02}Villani, C.: A review of mathematical topics in collisional
kinetic theory, in Handbook of Mathematical Fluid Dynamics, Vol. I,
Friedlander, S., Serre, D. eds., Chapter 2 (2002). https://www.sciencedirect.com/science/handbooks/18745792/1.

\bibitem{AV02}Alexandre, R., Villani, C.: On the Boltzmann equation
for long-range interactions, Commun. Pure Appl. Math. \textbf{55},
30--70 (2002). https://doi.org/10.1002/cpa.10012.

\bibitem{MS07}Mouhot, C., Strain, R. M.: Spectral gap and coercivity
estimates for linearized Boltzmann collision operators without angular
cutoff, J. Math. Pures Appl. \textbf{87}, 515--535 (2007). https://doi.org/10.1016/j.matpur.2007.03.003.

\bibitem{AMUXY10}Alexandre, R., Morimoto, Y., Ukai, S., Xu, C.-J.,
Yang, T.: Regularizing effect and local existence for the non-cutoff
Boltzmann equation, Arch. Rational Mech. Anal. \textbf{198}, 39--123
(2010). https://doi.org/10.1007/s00205-010-0290-1.

\bibitem{AMUXY11}Alexandre, R., Morimoto, Y., Ukai, S., Xu, C.-J.,
Yang, T.: Global existence and full regularity of the Boltzmann equation
without angular cutoff, Commun. Math. Phys. \textbf{304}, 513--581
(2011). https://doi.org/10.1007/s00220-011-1242-9

\bibitem{GS11}Gressman, P. T., Strain, R. M.: Global classical solutions
of the Boltzmann equation without angular cut-off, J. American Math.
Soc. \textbf{24}, 771--847 (2011). https://doi.org/10.1090/S0894-0347-2011-00697-8.

\bibitem{CH11} Chen, Y., He, L.-B.: Smoothing estimates for Boltzmann equation with full-range interactions: Spatially homogeneous case, 
Arch. Rational Mech. Anal. \textbf{201}, 501--548 (2011). doi: 10.1007/s00205-010-0393-8

\bibitem{JL19}Jiang, J.-C., Liu, T.-P.: Boltzmann collision operator for the infinite range potential: A limit problem, Ann. I. H. Poincar\'{e} \textbf{36}, 1639--1677 (2019). https://doi.org/10.1016/j.anihpc.2019.03.001


\bibitem{T15}Takata, S.: A toy-model study of the grazing collisions
in the kinetic theory, J. Stat. Phys. \textbf{160}, 770--792 (2015).
https://doi.org/10.1007/s10955-015-1259-0.

\bibitem{B94}Bird, G. A.: Molecular Gas Dynamics and the Direct
Simulation of Gas Flows, Clarendon Press, Oxford (1994).

\bibitem{SOA89}Sone, Y., Ohwada, T., Aoki, K.: Temperature jump and
Knudsen layer in a rarefied gas over a plane wall: Numerical analysis
of the linearized Boltzmann equation for hard-sphere molecules, Phys.
Fluids A \textbf{1}, 363--370 (1989). https://doi.org/10.1063/1.857457.

\bibitem{HT15}Hattori, M., Takata, S.: Second-order Knudsen-layer
analysis for the generalized slip-flow theory I, Bulletin of the Institute
of Mathematics, Academia Sinica (New Series) \textbf{10}, 423--448
(2015).

\bibitem{KR19}Kessler, T., Rjasanow, S.: Fully conservative spectral
Galerkin--Petrov method for the inhomogeneous Boltzmann equation,
Kinetic \& Related Models \textbf{12}, 507--549 (2019). doi: 10.3934/krm.2019021.

\bibitem{SO73}Sone, Y., Onishi, Y.: Kinetic theory of evaporation
and condensation, J. Phys. Soc. Jpn \textbf{35}, 1773--1776 (1973)
https://doi.org/10.1143/JPSJ.35.1773; ibid, Kinetic theory of evaporation
and condensation---Hydrodynamic equation and slip boundary condition---,
J. Phys. Soc. Jpn \textbf{44}, 1981--1994 (1978). https://doi.org/10.1143/JPSJ.44.1981.

\bibitem{S64}Sone, Y.: Kinetic theory analysis of linearized Rayleigh problem, J. Phys. Soc. Jpn \textbf{19}, 1463--1473 (1964).
https://doi.org/10.1143/JPSJ.19.1463

\bibitem{K21}Koike, K.: Refined pointwise estimates for a 1D viscous
compressible flow and the long-time behavior of a point mass, RIMS
K\^oky\^uroku, Appendix~A.1 (to be published).

\bibitem{F}As of May 20, 2021, the library is available from http://www-an.acs.i.kyoto-u.ac.jp/\textasciitilde fujiwara/exflib/.

\bibitem{J}As of May 20, 2021, the library is available from https://fredrikj.net/python-flint/\#.
\end{thebibliography}
\end{document}